\documentclass[a4paper,11pt]{article}
\usepackage[utf8]{inputenc}
\usepackage{mathtools}
\usepackage{natbib}
\usepackage{graphicx}
\usepackage{adjustbox}
\usepackage[font=footnotesize]{caption,subcaption} 
\usepackage{array} 
\usepackage{listings} 
\usepackage{lineno} 
\usepackage[nottoc]{tocbibind} 
\usepackage{hyperref} 
\usepackage[percent]{overpic} 
\usepackage{float} 
\usepackage{chngcntr} 
\usepackage{color}
\usepackage{booktabs}
\usepackage{authblk}
\usepackage{fullpage}
\usepackage{amsmath,amssymb,dsfont}
\usepackage{amsthm}
\usepackage{placeins}
\usepackage{longtable}
\usepackage{xcolor}
\usepackage{eurosym}
\newcolumntype{L}[1]{>{\raggedleft\let\newline\\\arraybackslash\hspace{0pt}}m{#1}}
\newcolumntype{C}[1]{>{\centering\let\newline\\\arraybackslash\hspace{0pt}}m{#1}}
\usepackage{makecell} 
\usepackage{multirow}
\usepackage{verbatim}

\renewcommand*{\thefootnote}{\fnsymbol{footnote}}

\definecolor{lightgray}{gray}{0.9}

\begin{document}

\title{Floods do not sink prices, historical memory does: \\ How flood risk impacts the Italian housing market$^{\dagger}$}
\author[1,2]{Anna Bellaver$^{\star,}$}
\author[2]{Lorenzo Costantini$^{\star,}$}
\author[3]{Ariadna Fosch$^{\star,}$}
\author[4]{Anna Monticelli}
\author[2]{\hspace{50pt}David Scala}
\author[5]{Marco Pangallo$^{\ddagger,}$}

\affil[1]{Department of Computer, Control, and Management Engineering Antonio Ruberti, Sapienza University, Rome}
\affil[2]{Intesa Sanpaolo, Turin 10138, Italy}
\affil[3]{Institute for Biocomputation and Physics of Complex Systems, University of Zaragoza, Zaragoza 50018, Spain}
\affil[4]{Intesa Sanpaolo Innovation Center, Turin 10138, Italy}
\affil[5]{CENTAI Institute, Turin 10138, Italy}

\date{\today}

\maketitle
\footnotetext[2]{We would like to thank Paolo Bajardi, Diana Bruni, Matteo Coronese, Paola Danelutti, Adriano di Fabio, Barbara Lenti, Jacopo Lenti, Laura Li Puma, Michela Manuelli, Carla Monferrato, Yamir Moreno, Silvia Ronchiadin, Simone Scarsi, and Fabio Verachi for their helpful comments at various stages of this work, as well as seminar participants at EAERE and IAERE, where this work was selected among the top 5 presentations by young researchers, to be included in a special session at the Italian Economic Society 66th Annual Conference.}
\footnotetext[1]{Equal contribution.}
\footnotetext[3]{Corresponding author: marco.pangallo@gmail.com}

\renewcommand*{\thefootnote}{\arabic{footnote}}

\begin{abstract}
Do home prices incorporate flood risk in the immediate aftermath of specific flood events, or is it the repeated exposure over the years that plays a more significant role? We address this question through the first systematic study of the Italian housing market, which is an ideal case study because it is highly exposed to floods, though unevenly distributed across the national territory. Using a novel dataset containing about 550,000 mortgage-financed transactions between 2016 and 2024, as well as hedonic regressions and a difference-in-difference design, we find that: (i) specific floods do not significantly decrease home prices in areas at risk; (ii) the repeated exposure to floods in flood-prone areas leads to price reductions, up to 4\% in the most frequently flooded regions; (iii) low-income buyers are more likely to settle in flood-prone areas.
\end{abstract}
JEL Codes: D83, Q51, Q54, R31

\noindent Keywords: Flood risk; housing markets; hedonic regressions; awareness

\section{Introduction}

Research on the impact of flood risk on home prices, primarily based on U.S. data, typically finds small impacts, except in the years immediately following major flood events, when prices substantially decline in affected areas. For example, prior to Hurricane Sandy, flood risk had little influence on home prices in New York City. However, in the years after the storm, properties in flood zones --despite being undamaged-- sold for up to 8\% less than properties in safer areas~\citep{ortega2018rising}. \cite{bin2013changes} attribute this pattern to the availability heuristic~\citep{tversky1973availability}, suggesting that individuals must experience floods firsthand to perceive them as a significant threat and demand a price discount for homes at risk.

In this paper, we consider an alternative yet complementary hypothesis, which we term \textit{historical memory}. We suggest that, in places with storms not as destructive as hurricanes and different housing market institutions than in the U.S., isolated flood events may be insufficient to significantly shift market beliefs and prompt the pricing of flood risk properties. Instead, repeated flood exposure in the same region may have a more substantial influence. If this hypothesis is true, home prices in frequently flooded areas are likely to reflect flood risk more prominently than those in rarely flooded areas, even for properties that remain undamaged. 

We operationalize the concept of historical memory by constructing a measure of buyers’ \textit{awareness} of flood risk. This measure rises following each flood event but gradually declines as memories of past floods fade. A single flood, even if severe, may not raise awareness sufficiently for flood risk to become reflected in home prices, as buyers might perceive it as an isolated incident. In contrast, repeated exposure to flooding heightens awareness, making it more likely that flood risk is capitalized into property values. Our measure offers a concrete implementation of the ideas proposed by \cite{tobin1986theoretical} and \cite{pryce2011impact}, enabling empirical testing with real-world data.

To test this hypothesis, we conduct the first systematic study of the Italian housing market, which is an ideal case study for several reasons. First, Italy faces significant flood risk, with over 10\% of its population and buildings exposed to flood risk~\citep{trigila2021dissesto}, and approximately 40 major floods recorded since 2000~\citep{emdat}. Second, Italy recently experienced a major flood in the Emilia-Romagna region on May 16–17, 2023, which displaced 45,000 people and caused damages estimated at around 10 billion euros~\citep{emdat}. Third, flood exposure varies across regions: while flood risk affects the entire country, Northern Italy experiences more frequent events than the South. Fourth, unlike markets in the U.S. or the U.K., Italy lacks institutional requirements for flood risk disclosure during property sales and neither mandates nor promotes flood insurance. Finally, floods in Italy typically result from storms with a less catastrophic impact than the hurricanes or tropical storms common in the U.S. 

Given this environmental and institutional context, we frame and test our  historical memory hypothesis as follows. We argue that, without an institutional structure that encourages the sharing of information regarding flood risk, households rely on the historical memory of major floods to identify the most at-risk areas. Due to the relatively limited destructive potential of individual floods, a single event may not significantly shift market beliefs. However, repeated flood occurrences gradually reinforce risk awareness, eventually leading to its capitalization in home prices. The recent Emilia-Romagna flood provides an opportunity to test the first aspect of this hypothesis, while the variation in flood risk across the Italian territory enables us to examine the second.

The main reason why there is so little research on how flood risk impacts the Italian housing market is that systematic data only started to exist in the last few years.\footnote{We are only aware of one study assessing the impact of flood risk on prices~\citep{rosato2017assessing}, using 197 data points manually collected from real estate agencies in Venice. Only recently systematic data about the housing market became available from the housing advertisement website \textit{Immobiliare.it}~\citep{pangallo2018home,loberto2022online}, but concerning our research question, these data have only been used to study the impact of a sea wall in Venice \citep{benetton2023does} or floods in Emilia-Romagna \citep{gentili2025impact} (the study by \cite{gentili2025impact} is posterior to our work, which was first posted online in February 2025).} 
In this paper, we take advantage of a novel database that contains the near-universe of mortgages issued by Intesa Sanpaolo, the largest Italian bank. Our final dataset contains about 550,000 mortgages issued from 2016 to 2024, which is about 15\% of Italian mortgage-financed transactions, and 10\% of total transactions.\footnote{The data are highly representative of the Italian housing market throughout the entire national territory, see Supplementary Section~\ref{sec:representativity}.} For each transaction, we know the sale price, the date, the income of the buyer, a proxy for whether the buyer is under-36, and several physical characteristics of the transacted home, including its geographical coordinates. We use the coordinates to determine whether a home is inside a flood-prone area by performing spatial matching with risk maps provided by ISPRA, the Italian Institute for Environmental Protection and Research~\citep{trigila2021dissesto}. Using the same approach, we also assess whether homes have been \textit{hit} by specific flood events, such as the 2023 Emilia-Romagna flood. By ``hit'', we mean that their coordinates lie within the maximal flooding area identified by the Copernicus EU program~\citep{copernicus}. 

First, we assess whether specific flood events lead to stronger price penalties in the affected areas. Previous studies, mostly based on the U.S., find price reductions in a 5-25\% range, that only last a few years after the event, and interpret these findings as evidence for a transitory increase in flood risk awareness.\footnote{See \cite{hallstrom2005market,daniel2009floods,kousky2010learning,atreya2013forgetting,bin2013changes,atreya2015seeing,ortega2018rising,beltran2019impact}.} In line with most of the literature, we follow a difference-in-difference approach that distinguishes between homes at risk and homes not at risk, as well as homes sold before or after the event. In addition to most of the literature, we further distinguish between homes that were hit by the flood and homes that were not hit but could have been, as they were in risky areas in the same municipalities as hit homes~\citep{ortega2018rising}. 

We find that the 2023 Emilia-Romagna flood (which, as discussed, has been the largest flood in Italy in the last 20 years) did not lead to any further price reduction for homes at risk that were not hit.\footnote{We also find that prices of hit homes decreased by 10\% in the first 9 months after the event, reversing to a 10\% appreciation between 9 and 15 months after the event. Most likely, affected homes were first damaged and then renovated, resulting in a price premium after renovations ended.} This result, which contrasts with much of the literature, is robust across specifications, home types (e.g., focusing on ground floor apartments), and considering the second and third largest flood events in the studied period.\footnote{This result is also independently confirmed by \cite{gentili2025impact}, who used a different dataset based on the universe of postings on Immobiliare.it \citep{loberto2022online}.   }

Second, we assess whether flood risk reduces prices, independently of specific flood events, and how this varies across parts of Italy with different levels of exposure. All else equal, economic theory predicts a price penalty for homes at risk~\citep{bin2013changes}. However, because of confounders, such as water-related amenities~\citep{bin2006real,daniel2009flooding}, it is hard to establish causality in empirical studies,\footnote{For instance, nice views over rivers or lakes may both increase prices and be correlated to high flood risk.} especially given that we cannot rely on repeated sales~\citep{beltran2019impact} or changes in risk maps~\citep{hino2021effect}.
To address this issue, we run hedonic regressions that take into account multiple physical characteristics of transacted homes, as well as granular spatial fixed effects, down to the census tract level. We argue that these fixed effects capture enough spatial variation so that the coefficient on flood risk is likely to capture the actual penalty of being located in a flood zone. 

We find that homes at risk sell at approximately 1\% less than homes that are not at risk. This risk penalty is not uniform. Dividing all transactions in our dataset by our measure of flood risk awareness in the area and at the time of the transaction, we find that flood risk is capitalized in home prices only in the top awareness tercile, where discounts reach 1.7\%. The regions which averaged highest awareness over the considered period show stronger discounts, reaching 2.5\% in Liguria, 3.5\% in Emilia-Romagna and 4\% in Toscana. Disaggregating discounts by buyers’ socioeconomic characteristics, we further find that only high-income individuals obtain a price discount when purchasing properties at risk, while no comparable effect is observed for low-income buyers.

Third, we examine whether the income of buyers purchasing homes in flood-risk areas differs from that of buyers purchasing in safer locations. In line with \cite{zivin2023hurricanes}, we find that lower-income households are more likely to buy properties in risky areas. This effect is driven by experienced buyers, while it is absent for young buyers under 36.

Summing up, in the immediate aftermath of the largest Italian floods in the past 20 years, we do not observe any price reductions for homes located in affected municipalities but not directly hit by the event. At the same time, we find a small baseline penalty associated with flood risk, which increases by up to a factor of four in the most flood-prone parts of Italy. We interpret this evidence as supporting our hypothesis of \textit{historical memory}, whereby it is not isolated floods, but the repeated exposure to floods, that drives the incorporation of flood risk in home prices. In addition, we document income disparities both in the ability to obtain a discount and in the likelihood of settling in flood-risk areas, possibly reflecting differences in bargaining power, education, and financial constraints.

\paragraph{Relation to the literature.} The studies quantifying the effect of flood risk on home prices have been reviewed in~\cite{daniel2009flooding}, \cite{beltran2018flood} and~\cite{contat2024climate}. 

A first question is how much specific flood events heighten housing market participants' perceptions about climate risk, leading to stronger price penalties for homes located in risk zones. Lots of studies address this question~\citep{hallstrom2005market,daniel2009floods,kousky2010learning,atreya2013forgetting,bin2013changes,atreya2015seeing,ortega2018rising,beltran2019impact}, generally finding a temporary increase in the price penalty that only lasts few years.\footnote{\cite{zivin2023hurricanes} are among the few that find a price premium in the years following hurricanes, suggesting that it may be due to a supply shortage due to destroyed homes.} For instance, \cite{bin2013changes} argue that the availability heuristic~\citep{tversky1973availability} explains why individuals need to experience flood events to attach a non-negligible probability to their occurrence. They find that two hurricanes occurring in 1996 and 1999 in North Carolina, after no major hurricane since 1954, lead to price reductions of 5.7\% and 8.8\%, respectively. The fact that the second hurricane led to higher price reductions than the first supports our awareness measure. \cite{atreya2015seeing} argue that it is important to distinguish between inundated and non-inundated homes, because price penalties may conflate the physical damage that occurred to homes and the increased attention that buyers may pay to flood risk. \cite{ortega2018rising} are able to clearly distinguish between the two cases, showing that price penalties in New York homes damaged by Hurricane Sandy were up to 17-22\% for several years, while penalties for homes at risk that were not damaged were around 8\%. Our results also show a strong decrease in prices of affected homes, around 10\% between 3 and 9 months after the Emilia-Romagna flood, but reconstruction is much faster, as prices appear to recover already after 9 months (presumably, this is linked to the difference in intensity between Hurricane Sandy and the Emilia-Romagna floods). Differently from most of the literature, however, we do not find any significant change in the price penalty for homes that were not directly hit by the flood, but still located in risky areas of affected municipalities.\footnote{An exception to the general findings of the literature is \cite{kousky2010learning}, who do not find further reductions in 100-year floodplains in Missouri, as flood risk seemed to be already capitalized in home prices.}

The next question is how much flood risk is capitalized in home prices in ``normal times'', i.e. not in the immediate aftermath of specific flood events.\footnote{Several studies use spatial regressions to estimate these effects~\citep{atreya2013forgetting, bin2013changes,zhang2016flood,pommeranz2020spatial}. Here, we do not find relevant spatial correlations after controlling for geographical units, as evidenced by Global Moran's I tests and LISA cluster maps, so we did not follow this direction. See Supplementary Section~\ref{sec:spatial_correlation}.} Almost all previous studies relied on U.S. data. \cite{daniel2009flooding} review estimates that range from a -52\% penalty for at-risk homes to a +58\% premium (mostly due to failing to control for water-related amenities), with a mean estimate of -2.3\%. \cite{beltran2018assessing} update these estimates considering more studies, and find a range between a -75.5\% penalty and 61\% premium, with an adjusted mean -4.6\% penalty for inland flooding. More recently, in what is perhaps the most complete study on this question, \cite{hino2021effect} find smaller -2.1\% and -1.4\% penalties with panel data and diff-in-diff approaches, respectively. Using actuarially fair insurance costs, they argue that these discounts are much smaller than the ones that would be expected in efficient markets, which would instead range from -4.7\% to -10.6\%. Our average 1\% price penalty for homes located in flood zones relates to these findings. Unsurprisingly, we find a smaller penalty compared to the U.S., as in Italy disclosure of flood risk is non-existent and flood insurance is in many cases not even encouraged.\footnote{It is harder to say to what extent flood risk is capitalized in home prices, because although the penalty is smaller in Italy than in the U.S., the extent of damages is also likely to be smaller, as Italian storms are less destructive than hurricanes in the U.S.} As mentioned above, our estimates of price penalties depend on buyers' awareness of flood risk, operationalizing ideas put forward by \cite{tobin1986theoretical} and~\cite{pryce2011impact}. 

The third piece of literature that we relate to is the one looking at the socio-economic characteristics of the buyers. Although there is substantial evidence that ``sophisticated'' buyers, such as non-owner-occupiers \citep{bernstein2019disaster}, or ``believers'' in climate change \citep{baldauf2020does}, obtain higher discounts when purchasing at-risk properties, the evidence on the effect of age or income is more scarce. For instance, \cite{bradt2021voluntary} find that younger and lower-income buyers are less likely to purchase flood insurance, but \cite{fairweather2024expecting} do not find any significant difference in terms of searching at-risk properties across income levels. We contribute to this literature by showing interactions between the age and income of the buyers. Regarding the income of buyers who settle in at-risk areas, our main comparison is \cite{zivin2023hurricanes}, who match information on buyers' income with transactions. They show that periods in which prices increase correspond to periods in which higher-income buyers are more likely to purchase properties in affected areas. Our results are closest to \cite{ellen2024heterogeneity}, who find that lower-income households moved into the areas of New York that were hit harder by Hurricane Sandy.

\paragraph{Roadmap.}
First, Section~\ref{sec:data} describes the data from Intesa Sanpaolo, the data from ISPRA and Copernicus spatially matched to transacted homes, as well as EM-DAT dataset used to quantify our awareness measure. Next, Section~\ref{sec:main_estimates} describes our econometric approach. The main results are in Section~\ref{sec:results}. Section~\ref{sec:conclusion} concludes.

\section{Data description}
\label{sec:data}

\subsection{Intesa Sanpaolo mortgages}
\label{sec:isp_mortgages}

\textit{Intesa Sanpaolo} (ISP) is the largest banking group in Italy, financing about a quarter of all mortgages within the country.\footnote{\url{https://www.agipress.it/mutui-vassena-intesa-sanpaolo-leader-in-italia-con-il-26-del-mercato}} Following a GDPR-compliant and privacy-preserving protocol, we got access to the vast majority of the mortgages that they issued from 2016 on. Our dataset contains all mortgages that are still being repaid at the time of data extraction, that is August 2024.\footnote{This excludes all the mortgages that were issued but have already been fully repaid, and mortgages that were issued but have been subrogated. However, focusing on mortgages that are less than 8 years old, these cases are not very common and we estimate that even in the initial year of the sample (2016-2018) we have 80\% of the mortgages that were actually issued.} The raw dataset contains 798,237 rows. We filtered data by selecting mortgages used to finance purchases (rather than, for instance, renovations), with available coordinates and within the Italian territory\footnote{Homes without available coordinates had to be removed from the analysis due to the impossibility of classifying them into risk areas.}, and following a few other criteria.\footnote{We also discarded the mortgages that were given to juridical persons, that are still under review, that were issued to construct housing units with the goal of reselling them, and that were acquired through an auction process.} The final dataset has 552,761 rows. 

ISP provided us with two databases: \texttt{Contracts} and \texttt{Cadaster}. The first one contains information about the mortgage applicant, the type, amount and goal of the mortgage, the issuance date and the price of the associated transaction. It also contains some information on the main housing unit given as collateral. In this database, a unique entry is a mortgage contract. The \texttt{Cadaster} dataset contains information about each cadastral unit: cadastral code (corresponding to apartment, detached house, garage, basement, etc.), size, energy class, position, floor and others. Here, a unique entry is a cadastral unit given as mortgage collateral.\footnote{In principle, mortgage applicants could give as collateral a housing unit different from the one they are buying, and we cannot tell if this is the case from our data. However, ISP states that these cases are residual for the subset of the data that we selected. For instance, it would be standard to give a home in Italy as collateral to buy a home abroad, but we exclude these transactions from our sample.}

We processed and cleaned these data, to create a single database where each sample is a mortgage contract, and the linked housing unit combines the information from all cadastral units given as collateral. To do so, we summed the floor areas of the \textit{residential} cadastral units that were given as collateral for the mortgage. This excludes, for instance, the basement or garage of a building, but it sums the floor area of, for example, two apartments that were independent cadastral units at the time of purchase but then are joined following renovation work. Another example that required data processing is the floor variable. Because some housing units are divided among multiple floors, we created a flag that takes value one if it is on multiple floors, and a numeric variable that takes the minimum floor in case of multi-floor apartments.\footnote{This is because the minimum floor is the one most at risk of flood.} More details on the data and the cleaning process are given in Supplementary Section~\ref{sec:datadescription}.

The ISP dataset contains the coordinates (latitude and longitude) of the housing units given as collateral for each mortgage.
This information is necessary to identify the risk level of the housing units (Section \ref{sec:ispra}), and to associate homes to specific spatial units such as municipalities, OMI microzones or census tracts.

\subsubsection{Summary statistics}

\begin{table}
\centering
\caption{Summary statistics of numeric variables. p1, p25, p75 and p99 refer to the 1st, 25th, 75th and 99th percentiles.}\label{tab:summary_statistics_numeric}
\begin{adjustbox}{width=1.1\textwidth, center}
\begin{tabular}{lrllllll}
\toprule
Variable & Missing & p1 & p25 & Median & Mean & p75 & p99 \\
\midrule
Issuance Date & 0 (0\%) & 2016-09-21 & 2019-03-22 & 2021-04-14 & 2020-12-23 & 2022-08-04 & 2024-07-19 \\
Construction Year & 1192 (0.22\%) & 1900 & 1965 & 1974 & 1979 & 2002 & 2022 \\
Surface Area (m$^2$) & 0 (0\%) & 42.00 & 83.00 & 107.00 & 121.57 & 143.00 & 345.00 \\
Sale Price (\euro) & 0 (0\%) & 44,000 & 110,000 & 155,000 & 184,510 & 225,000 & 660,000 \\
Floor (numeric) & 447545 (80.97\%) & 0.00 & 0.00 & 1.00 & 1.65 & 2.00 & 8.00 \\
Monthly Income (\euro) & 0 (0\%) & 873 & 1,670 & 2,374 & 2,870 & 3,413 & 10,700 \\
Latitude  & 0 (0\%) & 37.51 & 41.85 & 44.38 & 43.54 & 45.48 & 46.14 \\
Longitude  & 0 (0\%) & 7.50 & 9.21 & 11.35 & 11.53 & 13.00 & 18.03 \\
\bottomrule
\end{tabular}
\end{adjustbox}
\end{table}

Table~\ref{tab:summary_statistics_numeric} shows summary statistics for numeric variables. We show the count of valid values, the count of missing values, the 1st and 25th percentiles, the median, the mean, and the 75th and 99th percentiles. On issuance date, we see that the data are evenly distributed from 2016 to 2024. Next, we see that 75\% of homes were constructed after 1965, and 50\% of homes have a surface area between 83 and 143 m$^2$. The median sale price is 155,000\euro. The vast majority of apartments are on the ground, first or second floor. The monthly income of the mortgage applicant is typically between about 1,500\euro \hspace{3pt} and 3,500\euro. 

Table~\ref{tab:summary_statistics_categorical} provides summary statistics for categorical variables. The share of mortgages generally reflects city size. As expected, the most common cadastral codes correspond to residence types, which refer to the main housing units used as collateral. Among these, standard (A02) and low-cost (A03) residences account for approximately 85\% of the transactions. 7\% of homes for which floor information is available span multiple floors, typically the ground and first floors. Additionally, about 50\% of homes have a garage, and 15\% include an additional unit, such as a basement or attic.
Regarding energy efficiency, most homes fall into the lowest energy classes: G, F, and E (about 48\% across these three energy classes). Finally, about one in four homes (among the ones that report this information) is equipped with air conditioning.

\subsubsection{Representativity}

Since this is the first time that ISP mortgage data are used for research, we need to evaluate how representative they are of the Italian housing market. We show in Supplementary Section~\ref{sec:representativity} that they are highly representative. Since ISP has branches across the entire country, the share of transactions in each province matches very well the share of transactions in administrative data (Pearson correlation: 0.94). ISP data also generally match administrative data in terms of home size, although there are relatively fewer small apartments in ISP data because buyers are less likely to apply for a mortgage in this case. In terms of prices, ISP data also match administrative data very well. For instance, when focusing only on the provincial capitals, we obtain a Pearson correlation of 0.89 between ISP and administrative data. Finally, we also show that the incomes declared by ISP mortgage applicants match those of Italian residents across provinces (Pearson correlation: 0.82), although mortgage applicants' incomes are slightly higher than the general population. In sum, our results suggest that ISP data are representative of the Italian housing market.

\subsection{ISPRA risk maps}
\label{sec:ispra}

ISPRA is the Italian organization monitoring hydrological risk across Italy. Specifically, ISPRA collects and aggregates the risk estimations performed by all Italian river basin authorities, providing freely available hydrological risk maps at the national scale.\footnote{\url{https://idrogeo.isprambiente.it/app/page/open-data}} 
According to these maps, 14\% of Italian territory is considered at flood risk. Zooming in at the region scale (i.e., at the sub-national level), the regions showing a fraction of territory at flood risk larger than the national average are Lombardia, Veneto, Friuli-Venezia-Giulia, Emilia-Romagna, Toscana and Calabria~\citep{trigila2021dissesto,iadanza2021idrogeo}.

In this work, we used the 2020 release of the ISPRA maps, which is the most recent update, as a reference for flood risk by the Italian authorities. The most recent tools for hydrological modeling and the inclusion of areas not considered in previous releases were adopted in the 2020 release. Despite these improvements, there are still some limitations in the definition of risk areas due to the uncoordinated work of the basin authorities. The main limitation is the inconsistency between the \textit{level} of risk. Although in principle ISPRA distinguishes between high, medium, and low risk, classified according to return time, this classification is inconsistent across regions.\footnote{\label{ft2}Italian law defines flood hazard classes based on return times (i.e., the inverse of the average frequency of occurrence): high hazard corresponds to return times between 20 and 50 years; medium hazard to return times between 100 and 200 years; and low hazard to return times above 200 years \citep{trigila2021dissesto}. In practice, however, classifications are not harmonized across river basin authorities, even within the same region: many authorities classify as ``medium hazard'' any area with a return time between 50 and 100 years (\url{https://www.isprambiente.gov.it/pre_meteo/file/Relazioni_metodologiche_mappe_II_ciclo.zip}). Additional inconsistencies arise from missing high-hazard data for the Marche region in the 2020 release \citep{trigila2021dissesto} and from the unusually large extent of high-hazard areas in Calabria, which results from the use of a buffer zone around the hydrographic network rather than a pure return-time criterion \citep{trigila2021dissesto}.}
Therefore, we will mostly rely on a generic flood risk \textit{indicator} that does not distinguish between risk levels. Throughout this work, we introduce the label ``at risk'' when a home is inside a risk area, without distinction across high, medium, and low levels. 

To match housing units to ISPRA maps, we perform a spatial matching approach, using the coordinates of each housing unit to match it to risk zones. From the summary statistics shown in Table~\ref{tab:summary_statistics_categorical}, we note that approximately 77\% of homes are at no risk of flooding, 11\% have low risk, 8\% have medium risk, and 4\% are at high risk (thus, 23\% of homes are in a flood-risk area). 

\subsection{Copernicus flooded areas}

The Copernicus Emergency Management Service provides geospatial information to support the management of natural disasters, like landslides, wildfires or floods~\citep{copernicus}.
For flooding events, this data includes detailed vector maps that outline the extent of the flooded area at different times during the emergency. 

We used the Copernicus flood maps to identify homes in the ISP dataset that could have been affected by the largest flooding events in Italy of the last years, such as the Emilia-Romagna floods of May 2023. Homes were matched to the flooded areas through a spatial matching approach. A home is considered as \textit{hit} by the flood if its geo-coordinates are located inside the maximal area affected by that flood event, regardless of the magnitude of the damages, as more detailed information is not available.\footnote{This spatial matching procedure relies on the exact geo-coordinates of homes. Given the sensitive nature of these data, we followed a privacy-preserving protocol that guarantees the confidentiality of ISP clients.} 

Table~\ref{tab:flood_events} shows the number of hit homes for all Copernicus events that are sufficiently large to be included in EM-DAT (see Section~\ref{sec:frequency} below). We see that the May 2023 Emilia-Romagna floods affected 1464 homes in our dataset, more than 90\% of all homes affected by any other considered flood. Therefore, we use this flood as a case study, and the second and third largest floods (in Toscana and Piemonte-Liguria, respectively) as robustness tests.

\subsection{Flood risk awareness: frequency of floods across Italy}
\label{sec:frequency}
To model buyers’ awareness of flood risk, we combined historical data on flood events in Italian regions with theoretical insights from the literature on risk perception. 

Historical data on flood events come from EM-DAT, a disaster database that records all floods occurring in Italy between 2000 and 2023\footnote{\label{ft1}Although EM-DAT includes earlier records, it advises against using pre-2000 data due to reliability concerns.}. Events are included if they satisfy at least one of the following criteria:
10 fatalities;
100 affected people;
a declaration of state of emergency;
a call for international assistance. 
The objectivity of these inclusion criteria, together with the widespread use of the database, makes EM-DAT an ideal starting point for our analysis. As a robustness check, we also consider an alternative dataset that we compiled from ISPRA data (see Supplementary Section~\ref{supp:ER_extra} for details). In both datasets, geographic information is provided at the regional level, making regions the natural spatial unit for computing awareness over time.

The theoretical literature on flood-risk awareness shows that risk perception typically rises after a major event and then gradually declines as the event fades from memory~\citep{pryce2011impact, tobin1986theoretical}. Several studies also estimate the time scale of this decay. For example, \cite{heimerl2014vorsorgender} and \cite{egli2002non} report that awareness halves after 7/10 years respectively, while \cite{barendrecht2019value} find a half-life of 17 years (i.e., the time required for the perceived risk to diminish by half).

We integrate these theoretical insights with the actual record of past flood events from EM-DAT to construct an estimate of flood awareness for each Italian region. (We cannot work at a finer spatial scale because information at the province level is often incomplete or inconsistent.) We define the awareness of flood risk in region $r$ on day $t$ as
\begin{equation}
a(r,t) = \sum_{d=t_0}^{t} 2^{-\frac{(t-d)}{\tau}} \cdot \mathcal{I}(r,d),
\label{eqn:awareness}
\end{equation}
where the indicator variable $\mathcal{I}(r,d)=1$ if EM-DAT reports a flood in region $r$ on day $d$ (and $0$ otherwise),\footnote{If a flood spans multiple days, we take the first day as $d$.} and $\tau$ is the \textit{half-life} parameter, representing the number of days required for the influence of a flood event to halve. Thus, the awareness in a region at time $t$ is a weighted sum of the flood events it has experienced, with each event’s contribution decaying exponentially as the time since the event increases.\footnote{We set $t_0=$ 01/01/2000 for the reasons discussed in Footnote~\ref{ft1}.} In this way, recent floods matter more, but their influence diminishes steadily as they are forgotten.

Overall, $a(r,t)$ combines in a single variable both the number of floods a region has experienced (\textit{spatial information}) and the time elapsed between each flood and the sale date (\textit{temporal information}), yielding a simple measure of buyers' awareness of flood risk.

Our measure is intentionally as assumption-free as possible, with $\tau$ as the only degree of freedom. All our analyses are robust across the three values $\tau = 7, 10, 17$ years reported in the literature \citep{heimerl2014vorsorgender,egli2002non,barendrecht2019value}. (We report results for the intermediate value $\tau=10$ in the main text and provide results for the other values in the Appendix.) Another advantage of this formulation is that it captures \textit{awareness} rather than \textit{realized damages}: it applies to the entire regional territory (most of which has not experienced floods), it spans a long pre-sample period (2000--2016), and it does not rely on event severity, which is often inconsistently recorded and difficult to match to regions in our data.

\begin{figure}[!h]
    \centering
    \includegraphics[width=0.75\textwidth]{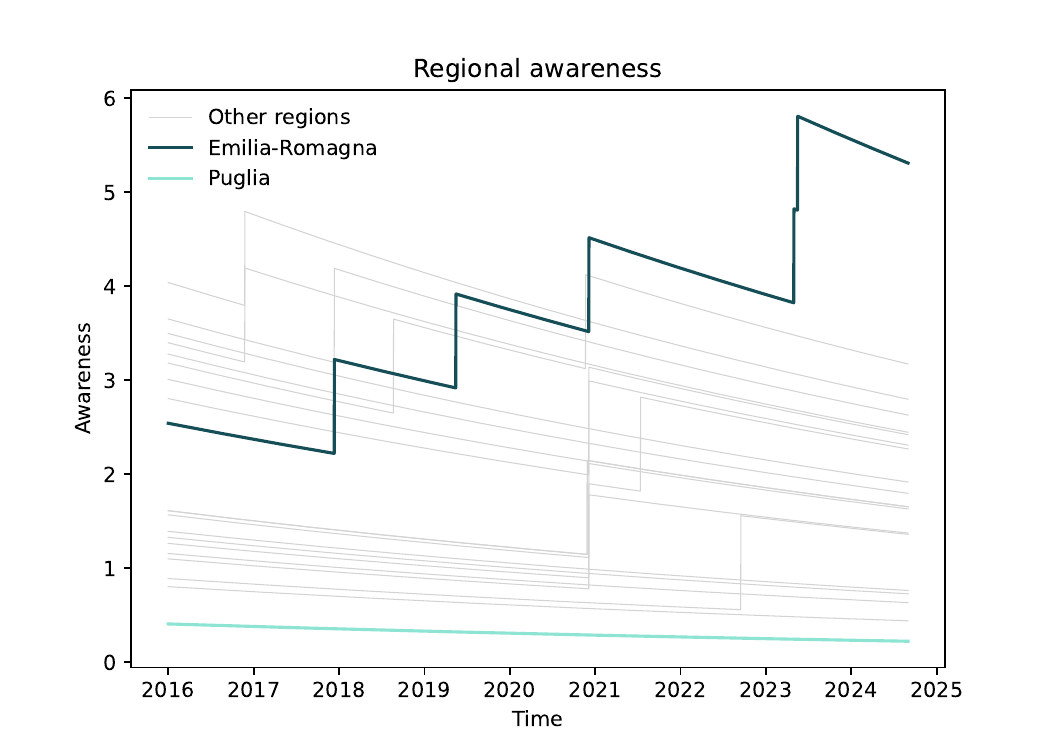}
    \caption{Regional awareness trends computed on the EM-DAT data. The awareness is computed day-by-day according to Equation~\eqref{eqn:awareness} with $\tau=10$ years. The spikes are due to flood events.}
    \label{fig:events_emdat}
\end{figure}

As an illustration, Figure~\ref{fig:events_emdat} shows the estimated awareness trends by region for the period covered by the ISP data (January 1st, 2016 to August 2024). Regions start the sample with different awareness levels due to their varying exposure to floods between 2000 and 2016. For example, Emilia-Romagna shows an awareness level of about 2.5 in 2016 and experienced several floods up to August 2024 (spikes in 2018, mid-2019, 2021, and mid-2023), producing an upward trend. In contrast, Puglia experienced only one flood in 2003, resulting in a low awareness level that steadily declines over time due to the exponential decay defined in Equation~\ref{eqn:awareness}.

Building on Equation~\eqref{eqn:awareness}, we associate with each transaction in the ISP dataset the awareness level at the time of sale $t_s$. That is, $a(r_s,t_s)$ quantifies the level of exposure to past floods that the buyer of each home faced at the moment of purchase.

\section{Econometric design} \label{sec:main_estimates}
To quantify the influence of flood risk on the Italian housing market, we have addressed three separate research questions: (1) Do we observe changes in home prices after specific flood events? (2) Does flood risk affect home prices in general? Is the effect larger by repeated exposure to floods (awareness) or by  socioeconomic characteristics of the buyers? (3) Are the incomes of buyers purchasing properties in flood risk areas different from the income of buyers in areas that are not at risk? In the following, we describe how we address all these questions by building on a baseline hedonic regression.

\subsection{Baseline hedonic regression}
Our baseline approach for identifying the effect of flood risk on home prices is a standard hedonic regression with granular fixed effects. Our goal is to compare the prices of very similar properties that differ only on flood risk. Denoting a transaction by $(i, k, t)$, where $i$ identifies the property, $k$ the spatial unit it belongs to, and $t$ the time of the transaction, our baseline hedonic specification is
\begin{equation}
\log{P_{ikt}} = \alpha_k + \alpha_t + \beta R_i + \boldsymbol{\gamma}' \textbf{X}_i + \epsilon_{ikt},
\label{eq:baseline}
\end{equation}
where $P_{ikt}$ is the transaction price, $\alpha_k$ denotes spatial fixed effects, $\alpha_t$ denotes time fixed effects, $R_i$ is a dummy variable that indicates that a home is at flood risk according to ISPRA maps,\footnote{Although it is unclear to what extent ISPRA flood maps are known to prospective buyers (given that they are publicly available but, as discussed in the introduction, flood risk information is not formally disclosed in Italy) we assume that these maps reflect an intuitive notion of flood risk. For example, proximity to a river is an obvious indicator of flood exposure, and such features are indeed captured in the ISPRA maps. Therefore, we use ISPRA classifications as a proxy for areas that buyers are likely to perceive as at risk.} and $\textbf{X}_i$ is a vector of property characteristics. We now discuss each of these terms one by one.

In our formulation, $\alpha_k$ captures all time-invariant home price differences due to being located in spatial unit $k$. We consider different definitions of spatial units, at different spatial aggregation levels. The widest definition of $k$ is the municipality, with a median area of 22 km$^2$. Especially for large municipalities, it is unlikely that this fixed effect captures enough difference in location, so we consider two further narrower definitions of $\alpha_k$. These are the OMI microzone and the census tract, respectively featuring a median area of 1.5 km$^2$ and 0.03km$^2$.\footnote{The OMI microzone is the standard spatial unit of the \emph{Osservatorio del Mercato Immobiliare} (OMI), the housing market observatory of the Italian Tax Office. It roughly corresponds to a neighborhood, although in rural areas it may contain larger portions of territory. Instead, census tracts correspond to blocks in cities and to villages (``frazioni") in rural areas. We use the latest available census tracts, namely from the 2011 census.} The inclusion of the $\alpha_k$ term completely captures the spatial correlation existing in our data, see Supplementary Section~\ref{sec:spat_corr}. This confirms that our hedonic regression approach is adequate and spatial econometric approaches are not necessary. 

Moreover, the fixed effect $\alpha_t$ is meant to capture price trends not linked to flood risk. To accommodate for potentially diverging trends across Italian provinces, we build fixed effects by considering the interaction of year and province indicators. We also consider monthly fixed effects to take seasonality into account.

Our specification of controls is completed with the vector $\textbf{X}_i$ subsuming the home characteristics available in our dataset: floor, surface area, energy class, cadastral class, building year, and indicators for the presence of a garage, an annex and air conditioning.\footnote{We build missing value dummies for most of these variables, so as to include data with missing values in the regression.} 

As we control for most confounders relating to the temporal and spatial dimension as well as most home features, we argue that $\beta$ captures the effect on home prices of  being located in a flood prone area.

\begin{figure}[!h]
    \centering
    \includegraphics[width=\linewidth]{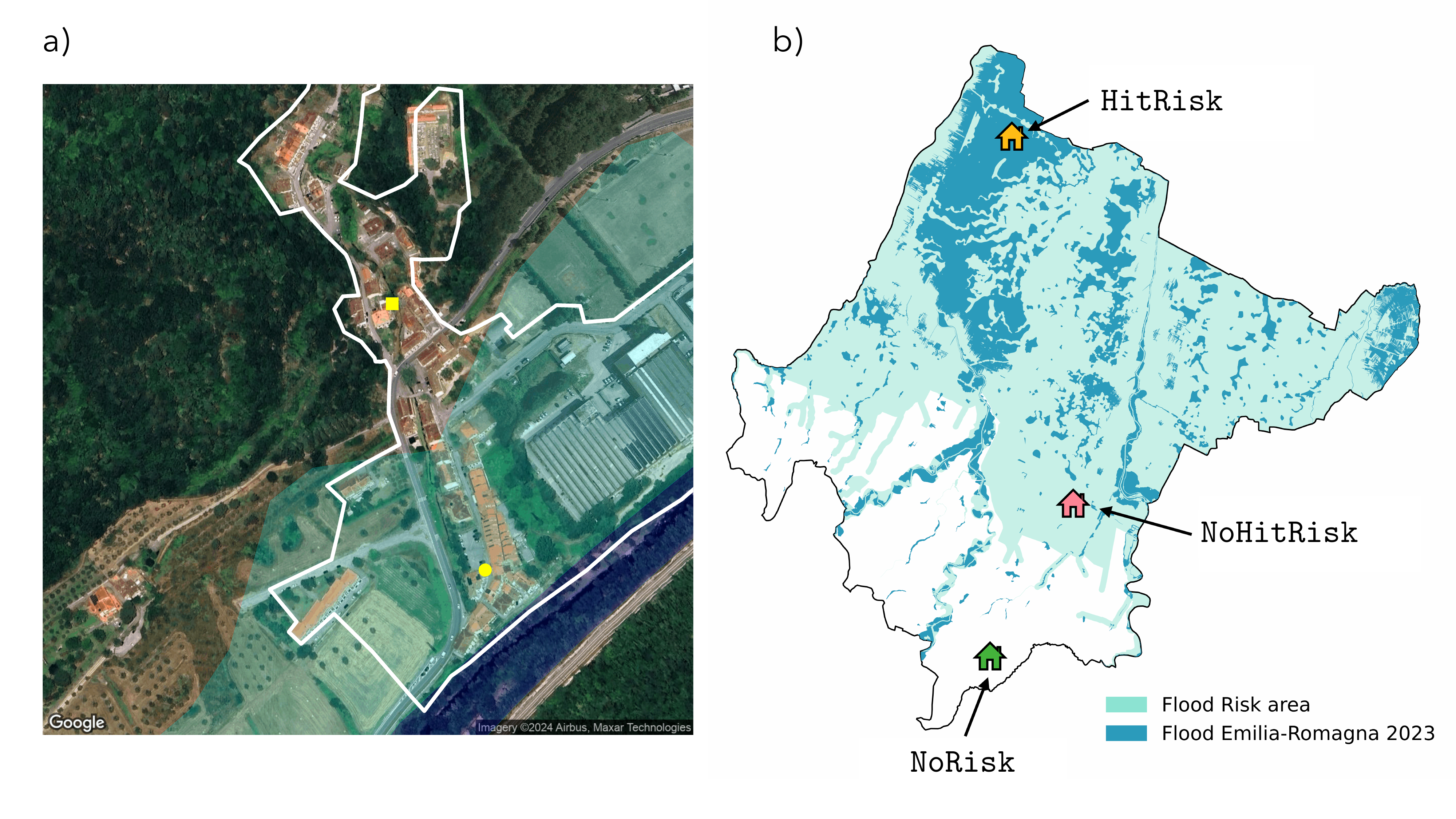}
    \caption{Graphical description of the regression framework. Panel a): An example to illustrate our identification strategy for homes at flood risk. We show two homes transacted in our dataset in yellow (after adding some noise to their positions to preserve privacy) in the village of Usella, Prato. We also show the boundaries of the census tract in white and the ISPRA risk areas in shades of blue (dark blue: high and medium risk, light blue: low risk). The home shown as a square at the top is not classified as at risk, while the other home shown as a circle is classified as at risk because it is close to the river. We argue that the residual difference in price between these two properties, after controlling for all other characteristics, is due to being located in the flood zone. Panel b): Graphical description of \texttt{HitRisk} and \texttt{NoHitRisk} homes considering the municipality of Forl\`i, one the cities affected by the Emilia-Romagna flood in 2023.}
    \label{fig:example_identification}
\end{figure}

Figure~\ref{fig:example_identification}a illustrates our identification strategy. As an example, we consider a village in the province of Prato, sided by the Bisenzio river. This river frequently floods, including most recently in November 2023, when it caused major damages to residential and industrial structures in the province of Prato (last line of Table~\ref{tab:flood_events}, Copernicus code EMSR705). Let us suppose that the characteristics of the two homes indicated in yellow are the same. Our spatial fixed effect, which in the most fine-grained specification corresponds to the census tract whose perimeter is shown in white, only includes the village. Thus, it is likely to absorb most differences in price for homes that are at most 500 meters from each other. This way, the coefficient on $R_i$ should only capture the exposition to flood risk.

\subsection{Effect of single flood events}\label{sec:diff_in_diff}
To estimate the impact that specific flood events had on home price (research question 1) we built on Equation~\eqref{eq:baseline} to formulate a hedonic difference-in-difference specification. Following~\cite{ortega2018rising}, we disaggregated the $R_i$ indicator into two categories, homes in a risky area that were hit by the flood (\texttt{HitRisk}) and homes that were not hit but are located in a risk zone of an affected municipality, and so can be considered a \textit{near-miss} (\texttt{NoHitRisk}).\footnote{We experimented alternative ways to characterize near-misses, including distance thresholds, with similar results. Our preferred assumption is the municipality because it does not require specifying a hard threshold for being considered a near-miss or not.} This distinction is exemplified in Figure~\ref{fig:example_identification}b.\footnote{Homes hit by the Emilia-Romagna flood (thus, sold after the flood), but not at risk (\texttt{HitNoRisk}) were not considered in the analyses because we have only 3 homes in this situation (we remind that the homes in the affected municipalities are 14977 and those hit by the flood are 1464, see Table~\ref{tab:flood_events}).} 

Since we aimed at estimating the price difference for homes \texttt{HitRisk} and \texttt{NoHitRisk} sold before and after a flood event (e.g. Emilia-Romagna flood of May 2023), we incorporated a temporal component ($\texttt{T}_i$) to differentiate between homes sold before and after the flood. Thus, our diff-in-diff specification considers the following model:

\begin{align}
\log{P_{ikt}} &= \alpha_k + \alpha_t + \beta_1\cdot\texttt{HitRisk}_i + \beta_2\cdot\texttt{NoHitRisk}_i  +  \nonumber \\
&\quad \delta_1 \cdot \texttt{T}_i \times \texttt{HitRisk}_i + \delta_2 \cdot \texttt{T}_i \times \texttt{NoHitRisk}_i + \boldsymbol{\gamma}' \textbf{X}_i + \epsilon_{ikt}.
\label{eq:diffindiff}
\end{align}
The categorical variable \texttt{T} reports the binned time of the transaction compared to the flood, using the period before the flood as a reference.
In this specification, the coefficients that we are mostly interested in are the ones on the interaction terms, namely $\delta_1$ and $\delta_2$, capturing, respectively, the additional reduction in prices for homes at risk in the flood affected area, and the additional reduction in prices for homes not directly hit by the flood but still considered at risk. In short, $\delta_1$ captures the direct effect due to being hit, whereas $\delta_2$ captures the indirect effect due to the beliefs of the market.

Finally, in Equation~\eqref{eq:diffindiff} the terms $\beta_1$ and $\beta_2$ capture the baseline risk for being located in a risk area. These two coefficients should thus be similar to one another (because the distinction between homes that were hit and not hit is already considered in the interactions) and should be similar to $\beta$ in Equation~\eqref{eq:baseline}.

\subsection{Disaggregated impacts of flood risk}\label{sec:hedonic_reg}

Our second research question aims at exploring the influence of flood risk ($R_i$) on home prices across all the national territory depending on different awareness conditions and buyers' characteristics. Our baseline equation, Equation~\eqref{eq:baseline}, is the starting point for this analysis. However, by interacting the indicator $R_i$ with other categorical/binary variables we can provide more insights about the heterogeneity beyond the global coefficient $\beta$.

\paragraph{Disaggregation by region.}
We disaggregated the effects of buying a home at risk across specific regions, providing a more granular overview of the heterogeneous impacts of flood risk across Italy. To do so we applied the following regression
\begin{equation}
\log{P_{ikt}} = \alpha_k + \alpha_t + \omega R_i \times \texttt{Region}_i + \boldsymbol{\gamma}' \textbf{X}_i + \epsilon_{ikt}.
\label{eq:region}
\end{equation}

\paragraph{Disaggregation by awareness.}
We also disaggregated the effects of buying a home at risk by buyers' flood awareness at the time of the transaction, i.e. our measure of historical memory derived from historical flood records (Section \ref{sec:frequency}). To this aim, we interact the risk indicator with a categorical indicator \texttt{Awareness} detailing the level of awareness in the region at the time of sale. As categories, we consider the terciles of the sale-time awareness distribution across all Italy, and name them \textit{high}, \textit{medium} and \textit{low} awareness\footnote{Categorical terms were introduced as dummy variables. Thus, when not interacted, we choose to remove  the \textit{low} awareness dummy to prevent colinearity.}. This results in the following equation
\begin{equation}
\log{P_{ikt}} = \alpha_k + \alpha_t + \rho R_i \times \texttt{Awareness}_i + \beta_{aw}\texttt{Awareness}_i+ \boldsymbol{\gamma}' \textbf{X}_i + \epsilon_{ikt}.
\label{eq:hist_mem}
\end{equation}

\paragraph{Disaggregation by socio-economic groups.}
\label{subsec:profile}

Although the theoretical foundations of hedonic regressions \citep{rosen1974hedonic} rely on competitive markets in which the marginal buyer alone determines prices, real housing markets are far from this ideal. Strong segmentation \citep{piazzesi2020segmented}, together with liquidity constraints and information asymmetries, allows for systematic price discrimination across different types of buyers. Following \cite{bernstein2019disaster}, we investigate whether buyers differ in their willingness to pay for flood risk depending on their age and income.

We encode the age of the buyer in the variable \texttt{Age} with two categories: \texttt{Young} and \texttt{NotYoung}. This is based on whether the mortgage was a \emph{mutuo giovani} (Young Buyers' Mortgage) or not. Transactions financed by the \emph{mutuo giovani} are associated with buyers under 36 with an adjusted annual income lower than 40,000\euro. We assumed all mortgages not labeled in the data as \emph{mutuo giovani} to refer to \texttt{NotYoung} buyers.\footnote{Note that only individuals under the age of 36 are eligible to apply for a \emph{mutuo giovani}. Conversely, while it is possible for a young buyer to obtain a standard mortgage, this is uncommon due to the more favorable terms offered by the \emph{mutuo giovani}.} 

For the level of income of the buyers, we divided buyers into 3 levels of income (high, medium, and low) based on the three terciles of the income distribution for each region. We define  the \texttt{IncomeLevel} of the buyers of each transaction by region because average income is widely different across Northern and Southern regions, and we want this variable to capture relative income within regions.

By interacting the $R_i$ indicator, \texttt{Awareness}, \texttt{Age}, and \texttt{IncomeLevel}, we can estimate whether young or more experienced buyers in different income classes obtain different price reductions when buying homes at risk across flood awareness levels. The equation used for this analysis is the following: 
\begin{align}
\log{P_{ikt}} =& \alpha_k + \alpha_t + \psi R_i \times \texttt{Awareness}_i \times \texttt{Age}_i \times \texttt{IncomeLevel}_i +\nonumber\\
&\quad \beta_{in}\texttt{IncomeLevel}_i + \beta_{age}\texttt{Age}_i + \beta_{aw}\texttt{Awareness}_i +  \boldsymbol{\gamma}' \textbf{X}_i + \epsilon_{ikt},
\label{eq:quadruple}
\end{align}
where the coefficient $\psi$ indicates the disaggregated effect of age, level of income, and flood awareness, on the price of homes at risk. Thus, it makes it possible to distinguish the reduction obtained by, e.g., low-income and high-income buyers that are both young.

\subsection{Income of the buyers moving to at-risk areas}
Beyond the evaluation of home prices in flood prone areas, we now focus on the income of people settling in risky areas. To explore this, we follow \cite{zivin2023hurricanes} and estimate the following regression: 
\begin{equation}
\log{I_{ikt}} = \alpha_k + \alpha_t + \beta R_i +  \boldsymbol{\gamma}' \textbf{X}_i + \epsilon_{ikt},
\label{eq:Incomebaseline}
\end{equation}
where $\log{I_{ikt}}$ is the log of the household's income and the other terms are equivalent to the ones described in Equation~\eqref{eq:baseline}.

Next, following the analyses on home prices, we zoom into the coefficient $\beta$ in Eq.~\eqref{eq:Incomebaseline} to explore whether flood-risk awareness and buyers' age play a role. For awareness, we change Eq.~\eqref{eq:hist_mem} to consider income ($I$) as dependent variable. For age, we consider the following specification:
\begin{align}
    \log{I_{ikt}} = &\alpha_k + \alpha_t + \phi R_i\times \texttt{Age}_i \times \texttt{Awareness}_i + \nonumber\\
    &\quad  \beta_{age}\texttt{Age}_i + \beta_{aw}\texttt{Awareness}_i + \boldsymbol{\gamma}' \textbf{X}_i + \epsilon_{ikt},
    \label{eq:incometriple}
\end{align}
where the coefficient $\phi$ quantifies whether young or experienced buyers purchasing homes at risk in certain awareness conditions have higher or lower income compared to similar people in areas that are not at risk.

\section{Results} \label{sec:results}

\subsection{Single floods only affect home prices in flooded areas}\label{sec:res_dif_in_dif}

\begin{figure}[t]
    \centering
    \includegraphics[width=\linewidth]{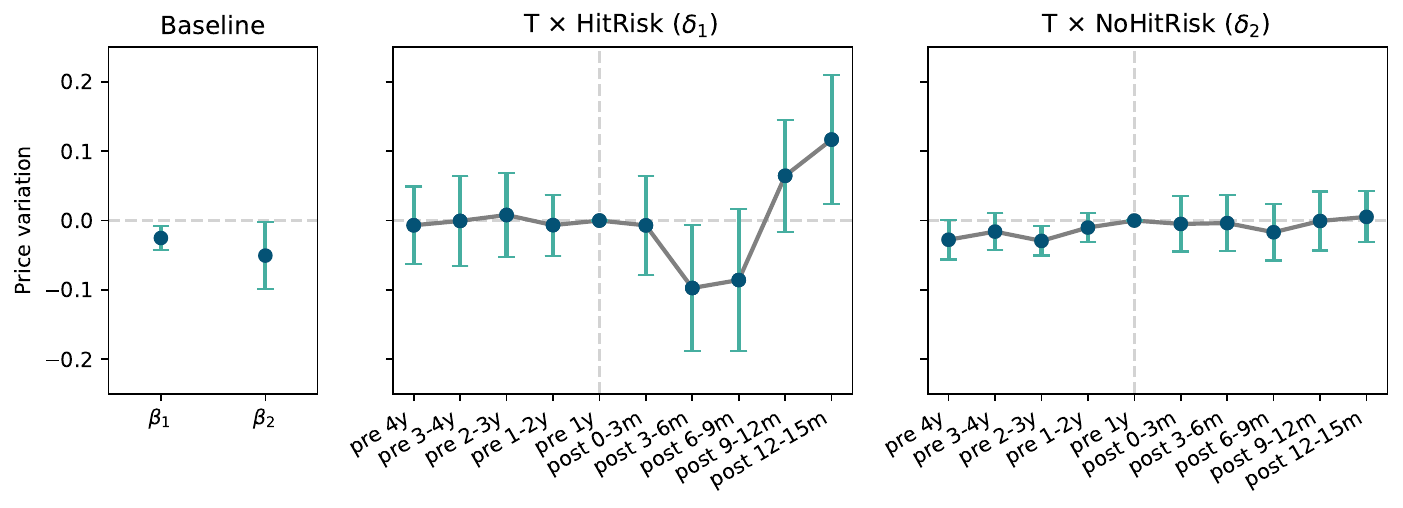}
    \caption{Time evolution of the home price variation for the 2023 Emilia-Romagna flood. Referring to Equation~\eqref{eq:diffindiff}, the left panel reports the coefficients $\beta_1$ and $\beta_2$, the central and right panels show the interaction coefficients $\delta_1$ and $\delta_2$ up to 15 months after the flood of Emilia-Romagna 2023. The temporal bins can be interpreted as follows. The labels \textit{pre} and \textit{post} identify whether a transaction has been completed before or after the considered flood event. The label \textit{y} indicates years and the label \textit{m} indicates months. We used the bin \textit{pre 1y} as a reference. Each point represents the average value of the regression coefficient and the error bars indicate the 95\% confidence intervals. We used spatial fixed effects at the OMI microzone level. R$^2$ = 0.74, observations = 35623.}
    \label{fig:quarters_EMSN154}
\end{figure}

To assess the impact that specific floods had on home prices we first focus on the major flood that hit Italy in the last 20 years, namely the May 2023 flood in Emilia-Romagna (coded EMSN154 by Copernicus). We later provide similar analyses for the 2016 Piemonte-Liguria (code EMSR192) and 2023 Toscana (EMSR705) floods in Supplementary Section~\ref{supp:ER_extra}.

Since the Emilia-Romagna flood is very recent, we only have observations up to 15 months after the flood event. We divide the year following the event in quarters and consider the interactions of quarters with \texttt{HitRisk} and \texttt{NoHitRisk} dummies (Equation~\ref{eq:diffindiff}). For instance, the interaction between a sale in the first quarter after the event and \texttt{HitRisk} indicates the reduction in price for homes that were hit by the flood and sold within 3 months of the event. 

Figure~\ref{fig:quarters_EMSN154} shows the results. The left panel shows the baseline discount for being in a risk zone for homes that were hit ($\beta_1$) and that were not hit ($\beta_2$). These coefficients indicate that \texttt{HitRisk} and \texttt{NoHitRisk} homes sell --on average-- around 3\% less than homes in safe areas, in line with results reported in the next subsection. 

Next, focusing on homes that were hit (central panel), we find no trend prior to the event, virtually no price reduction in the first 3 months after the event, and then a substantial reduction between 3 and 9 months from the event. In this period, homes sell at a further 10\% less than the baseline 3\% for homes at risk. However, in the last two quarters of observations, the prices actually increase compared to the baseline: homes that were hit by the flood sell at a 10\% premium compared to homes that were not hit (and not in the risk zone).\footnote{All the quarters post-flood have similar number of transactions, with an average of 363 transactions per quarter.} We interpret these results as follows. In the first three months after the event, prices did not decrease because many transactions reflect deals that were agreed upon before the flood (with possibly a pre-authorization on the mortgage amount by the bank). In the following months, the reductions in prices reflect the cost of the renovation work that falls on the buyer. Finally, 9 months after the event most renovation work has been carried out, and so the price increase reflects newly renovated homes.\footnote{We unfortunately do not have data on renovation status to test this hypothesis, but it could be tested in the future as ISP is increasingly collecting this information.} We think this is a reasonable timeline for renovation for the type of damage suffered by homes in this flood.

Turning to homes that were not hit, we find essentially no further reductions in prices after the event.\footnote{Before the event, the parallel trends assumption mostly holds except for the period between 2 and 3 years before the event, which however was the year of Covid-19, and this could bias the results. Furthermore, following~\cite{atreya2013forgetting} and~\cite{bin2013changes}, we also compared homes' hedonic characteristics one year before and after the event (Table~\ref{tab:statistic_test_pre_post}). We note a general agreement among the hedonic properties evaluated, especially referring to home size, location within a flood-risk area, cadastral code and the presence of garages and annexes.} This is our most surprising result, in contrast with much of the prior literature. It suggests that people were already discounting being located in a flood zone (homes at risk were selling at a 3.5\% discount even before the flood), and the actual event did not move their beliefs. 

Table~\ref{tab:statistic_test_pre_post} compares the statistical properties of the homes considered one year before and one year after this flood. These results show that the hedonic price function did not change after the flood event, supporting our interpretation of the coefficients. These results were also consistent when considering alternative parameter specifications (such as more granular fixed effects), when analyzing only ground-floor apartments or independent houses, and when exploring other major flood events (see Supplementary Section~\ref{supp:ER_extra}).\footnote{Our study on the 2023 Emilia-Romagna flood (EMS154) can be consistently compared to previous studies focusing on homes in the 100-year floodplain~\citep{atreya2013forgetting,bin2013changes}. In fact, almost all at-risk homes in the municipalities affected by the Emilia-Romagna flood are in high or medium risk areas; thus, they are characterized by return times less or equal than 200 years. We only find a negligible number of homes placed in the low-risk zones (9 homes in total). To the best of our knowledge, considering ISPRA data, high or medium risk areas are the those that better align with the 100-years flood zones used in previous studies. See also footnote \ref{ft2}.}

\subsection{Homes at risk sell at lower price under high awareness conditions} \label{sec:hedonic_risk}
In the left panel of Figure~\ref{fig:quarters_EMSN154}, we showed that being located in an area at risk leads to a lower home price, irrespective of the specific flood considered. To explore this question further, we estimate Equation~\eqref{eq:baseline} on our entire dataset aiming to provide a national-scale picture of flood risk across Italy.

\begin{table}[]
\caption{Flood risk effect on home prices. For each spatial fixed effects, Models 1-3-5 report the effect of the risk on home prices estimated using Equation~\eqref{eq:baseline}, while Models 2-4-6 disaggregate the effect of flood risk by the level of flood awareness at the time of the purchase, Equation~\eqref{eq:hist_mem}. Awareness is estimated following the procedure described in Section~\ref{sec:frequency} using the data reported by~\cite{emdat} and a half life time equal to 10 years. *** $p < 0.01$, ** $p < 0.05$, * $p < 0.1$.}
\begin{tabular}{l|cccccc}
\hline
 & \multicolumn{6}{c}{Dep. var.: log(\textit{P})} \\ \hline
\textbf{Model id} & 1 & \multicolumn{1}{c|}{2} & 3 & \multicolumn{1}{c|}{4} & 5 & 6 \\ \hline
\textbf{Risk} & \begin{tabular}[c]{@{}c@{}}-0.029$^{***}$\\ (0.010)\end{tabular} & \multicolumn{1}{c|}{} & \begin{tabular}[c]{@{}c@{}}-0.010$^{***}$\\ (0.003)\end{tabular} & \multicolumn{1}{c|}{} & \begin{tabular}[c]{@{}c@{}}-0.008$^{**}$\\ (0.003)\end{tabular} &  \\[15pt]

\textbf{\begin{tabular}[c]{@{}l@{}}Risk $\times$ High \\ awareness\end{tabular}} &  & \multicolumn{1}{c|}{\begin{tabular}[c]{@{}c@{}}-0.038$^{**}$\\ (0.016)\end{tabular}} &  & \multicolumn{1}{c|}{\begin{tabular}[c]{@{}c@{}}-0.017$^{***}$ \\ (0.004)\end{tabular}} &  & \begin{tabular}[c]{@{}c@{}}-0.012$^{***}$\\ (0.004)\end{tabular} \\[15pt]

\textbf{\begin{tabular}[c]{@{}l@{}}Risk $\times$ Medium \\ awareness\end{tabular}} &  & \multicolumn{1}{c|}{\begin{tabular}[c]{@{}c@{}}-0.031$^{***}$\\ (0.009)\end{tabular}} &  & \multicolumn{1}{c|}{\begin{tabular}[c]{@{}c@{}}-0.007$^{*}$\\ (0.004)\end{tabular}} &  & \begin{tabular}[c]{@{}c@{}}-0.008$^{*}$\\ (0.004) \end{tabular} \\[15pt]

\textbf{\begin{tabular}[c]{@{}l@{}}Risk $\times$ Low \\ awareness\end{tabular}} &  & \multicolumn{1}{c|}{\begin{tabular}[c]{@{}c@{}}-0.014$^{}$\\ (0.013)\end{tabular}} &  & \multicolumn{1}{c|}{\begin{tabular}[c]{@{}c@{}}-0.002$^{}$ \\ (0.004)\end{tabular}} &  & \begin{tabular}[c]{@{}c@{}}-0.005$^{}$\\ (0.004)\end{tabular} \\ \hline
\textbf{Observations} & \multicolumn{2}{c|}{552728} & \multicolumn{2}{c|}{552728} & \multicolumn{2}{c}{552728} \\
\textbf{R\textsuperscript{2}} & \multicolumn{2}{c|}{0.669} & \multicolumn{2}{c|}{0.755} & \multicolumn{2}{c}{0.848} \\ \hline
\textbf{Spatial fix. effects} & \multicolumn{2}{c|}{Municipality} & \multicolumn{2}{c|}{OMI zone} & \multicolumn{2}{c}{Census Tract} \\
\textbf{N. clusters} & \multicolumn{2}{c|}{7203} & \multicolumn{2}{c|}{20297} & \multicolumn{2}{c}{160630} \\
\textbf{Clustered s.e.} & \multicolumn{2}{c|}{\checkmark} & \multicolumn{2}{c|}{\checkmark} & \multicolumn{2}{c}{\checkmark} \\ \hline
\end{tabular}
\label{tab:NewBaselinePriceTable_EMDAT_numflood}
\end{table}

Table~\ref{tab:NewBaselinePriceTable_EMDAT_numflood} shows the results. We only report the effect of flood risk on price, but the other hedonic coefficients can be found in Table~\ref{tab:hedonic} (using Model 3 as a reference). In columns 1-3-5, we show the national average effect of flood risk on home prices, whereas columns 2-4-6 disaggregate this effect according to the level of awareness at the time of sale (high, medium or low awareness). Models 1-2 consider spatial fixed effects at the level of the municipality, while models 3-4 and 5-6 consider OMI microzone and census tract fixed effects, respectively. For each coefficient, we report the standard error and the significance level. In all cases, standard errors are clustered at the corresponding spatial level and we use the within estimator.\footnote{We tested a few specifications for the independent variables, such as whether we should consider the logarithm of the surface area, floor number as a numeric variable or binned categories, or whether we should remove outliers. In the end, we adopted the following baseline specification: logarithmic home size (m$^2$), binned home floor levels and removing the top and bottom 0.1\% samples for all numeric variables. Supplementary Section~\ref{sec:BaselineEq-SM} reports the robustness analysis for alternative specifications and further tests, like the robustness analyses across alternative half times ($\tau=7$ and $17$ years) and event records. All results shown are in line with those presented in the main text.}

Considering the national effect (first row), Model 1, which assumes municipality-level fixed effects, shows a 2.9\% price reduction for homes at risk, while Models 3 and 5, with OMI microzone and census tract fixed effects, show a 1\% and 0.8\% depreciation, respectively.\footnote{It was unexpected that larger aggregation levels for spatial fixed effects would lead to stronger price reduction. In fact, we expected the opposite assuming that there are positive amenities associated with being located close to rivers (such as water views). These amenities would lead to a positive effect of flood risk on home prices \citep{bin2006real,daniel2009flooding} when failed to be controlled for, such as when controlling for the municipality instead of more detailed OMI microzones or census tracts. Because we find a stronger reduction when controlling for the municipality, it is likely that in our sample there are on average no positive amenities for being located close to rivers. We think that this is reasonable given that we consider the entire national territory, including many rural areas.}\footnote{As described in Section~\ref{sec:ispra}, ISPRA divides flood-prone areas into low, medium, and high risk areas, but this classification is often inconsistent across regions. Table~\ref{tab:price_risk_levels} disaggregates the national effect of flood risk on prices, shown in the first row of Table~\ref{tab:NewBaselinePriceTable_EMDAT_numflood}, by risk level. In all cases, being at high risk has no statistically significant effect on prices, while being at medium or low risk has a negative and statistically significant effect on prices (but whether this effect is larger in medium or low risk areas depends on the considered spatial fixed effects). We interpret these results as evidence that the risk levels are not very informative due to idiosyncratic definitions by different river basin authorities, while an indicator of risk (irrespective of the risk level) is more likely to be informative. Further details can be found in Supplementary Section~\ref{sec:BaselineEq-SM}.}
Because results are very similar for OMI microzones and census tracts, from now on we consider OMI microzones as the baseline as they are the best trade-off between granularity and sample size.

Model 4 shows the disaggregation of flood risk effects when considering our awareness measure at the time of sale. Notably, the statistically significant 1\% price reduction at the national scale is mainly driven by the discount observed in high awareness conditions (-1.7\%). The other coefficients for flood risk are slightly or not statistically significant and show an average 0.7\% and 0.2\% price reduction for transactions with medium and low awareness. The leading role of high awareness transactions in determining the impacts of flood risk on prices is robust considering more granular fixed effects (Model 6 of Table~\ref{tab:NewBaselinePriceTable_EMDAT_numflood}), different event data and half-life parameter (we refer the reader to Supplementary Section~\ref{sec:BaselineEq-SM}, Tables~\ref{tab:SM_table2} and~\ref{tab:baselineRegr_ISPRA_numflood}, for further details).

\begin{figure}[ht]
    \centering
    \includegraphics[width=\linewidth]{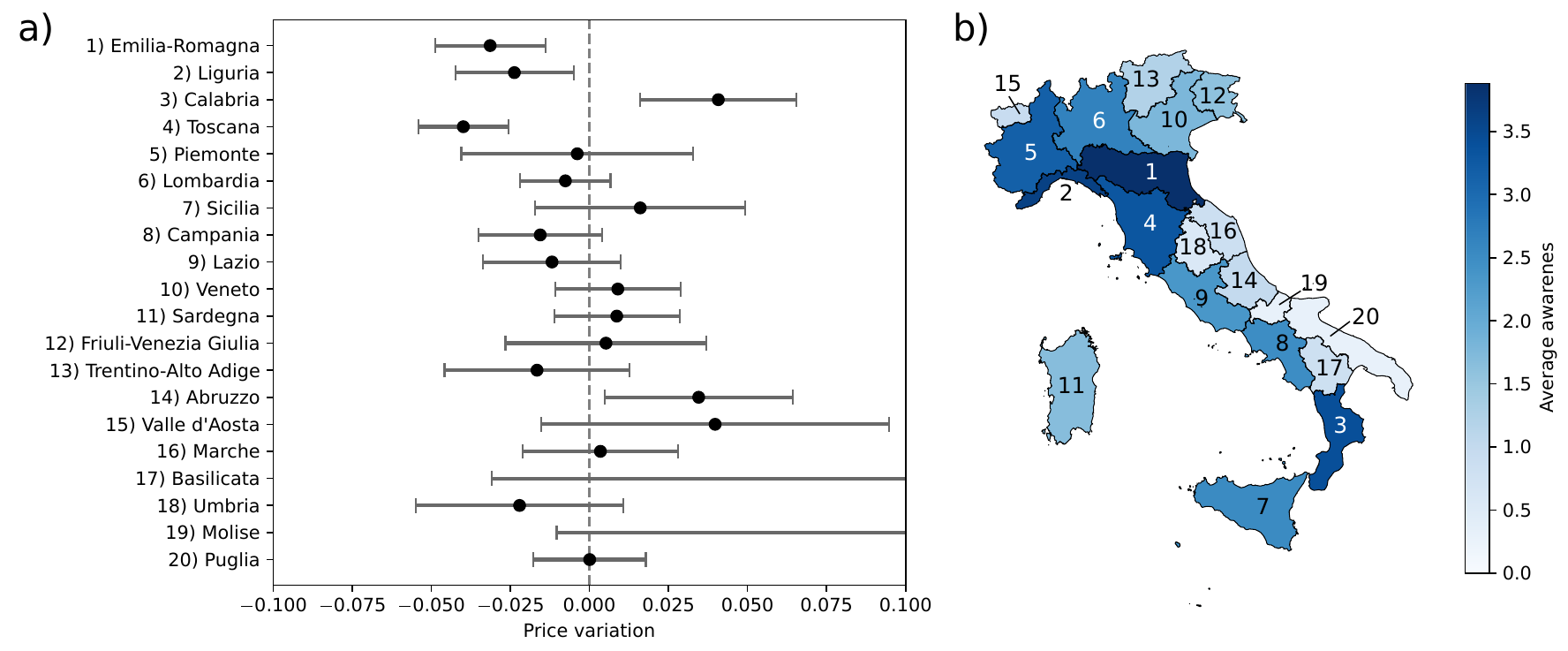}
    \caption{Effect of being at flood risk on home price, disaggregated by region. Panel a): The dots show the point estimate of the coefficients, while the error bars show 95\% confidence intervals. 
    For graphical purposes, we limited the x-axis between -0.1 and 0.1 (although not fully shown, the average price variations for Molise and Basilicata are 0.19 and 0.15, respectively). Regions are sorted according to time-averaged awareness: the first region (Emilia-Romagna) shows the largest average awareness value while the last region shows the lowest value. Panel b): Map of the Italian regions colored according to the average awareness at time of sale ($\tau=10$ years, see Section~\ref{sec:frequency}). Northern regions: Piemonte, Valle d'Aosta, Liguria, Lombardia, Trentino-Alto Adige, Veneto, Friuli-Venezia Giulia, and Emilia-Romagna. Center regions: Toscana, Umbria, Marche, Lazio, Abruzzo, and Molise. Southern and Islands regions: Campania, Puglia, Basilicata, Calabria, Sicilia, and Sardegna.}
    \label{fig:regions}
\end{figure}
To provide more insights into the spatial heterogeneity of flood risk we also disaggregated the national estimates across regions. Focusing on the same specification as Model 3 of Table~\ref{tab:NewBaselinePriceTable_EMDAT_numflood}, and interacting the risk indicator with region dummies, Figure~\ref{fig:regions} shows that the effect of risk on prices varies a lot across regions. In Emilia-Romagna, Liguria, and Toscana, homes at risk sell at a substantially higher discount than the national average (3.5\%, 2.5\% and 4\%, respectively, compared to 1\%), whereas in Abruzzo and Calabria they sell at a premium (especially for Calabria, this may reflect an idiosyncratic definition of risk in ISPRA maps, see Section~\ref{sec:ispra}).\footnote{Interestingly, the average flood risk coefficient for Emilia-Romagna agrees with $\beta_1$ and $\beta_2$ retrieved using the diff-in-diff set-up (Figure~\ref{fig:EmiliaRomagnaFlood_quarter}). These two coefficients do not need to correspond exactly to the coefficient on Emilia-Romagna in Figure~\ref{fig:regions} because they only refer to the affected municipalities and not to the entire region. However, their agreement reinforces our results both for the hedonic-regression and diff-in-diff analyses.} In most other regions the effects are not statistically significant at the 5\% level, although regions in the North and Center are more likely to show negative effects than regions in the South and Islands. When comparing these results with the average awareness at the time of sale for each region, we can observe some relevant correlations. Four out of five regions in the top 25\% awareness present price reduction due to flood risk (Emilia-Romagna, Liguria, Toscana, and Piemonte), while regions in the bottom 25\% exhibit price increments (Marche, Basilicata, Molise, and Puglia).

\subsubsection*{Higher-income buyers obtain stronger price reductions}

The results presented in the previous sections support our hypothesis that flood-risk-related price discounts are driven by the historical memory of past floods rather than by the occurrence of specific flood events. In this section, we investigate whether buyers with different socio-economic characteristics (specifically age and income) obtain heterogeneous discounts when purchasing homes exposed to flood risk.

\begin{figure}[!h]
    \centering
    \includegraphics[width=\linewidth]{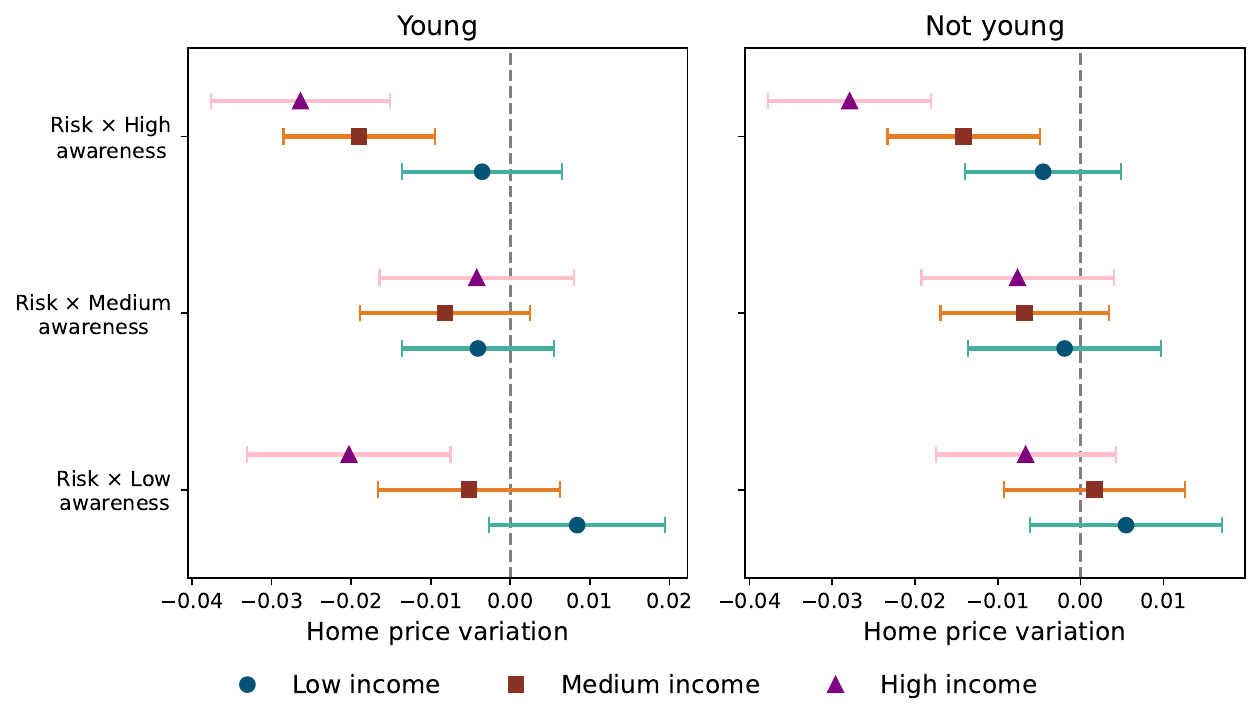}
    \caption{Flood risk effect on home prices disaggregated across awareness, mortgage applicant age, and income class of the buyer, Equation~\eqref{eq:quadruple}. The left panel shows the home price variation due to flood risk for transactions financed by the \emph{mutuo giovani}, and disaggregated by our classification of historical memory (i.e., awareness) and buyers' income class. The right panel reports the same analysis for transactions financed by a standard mortgage. The buyers' income class is estimated considering the top, middle and bottom terciles of the income distribution within the same region (see Section~\ref{subsec:profile} for further details).}
    \label{fig:QuadrupleInteraction}
\end{figure}

To this end, we estimate Eq.~\eqref{eq:quadruple} (see Section~\ref{subsec:profile}). Figure~\ref{fig:QuadrupleInteraction} reports the estimated flood-risk discount for \texttt{Young} and \texttt{NotYoung} buyers purchasing at-risk properties, distinguishing by buyers’ income levels and by the degree of flood-risk awareness (high, medium, and low). Under high-awareness conditions, \texttt{Young} (left panel) and \texttt{NotYoung} (right panel) buyers display remarkably similar patterns: high-income buyers obtain a flood-risk discount of approximately 3\%, while medium-income buyers receive a discount of around 2\%. As flood-risk awareness declines, the effect of flood exposure becomes statistically insignificant across most groups. The only notable exception is \texttt{Young} high-income buyers in low-awareness areas, who still obtain a discount of roughly 2\% when purchasing homes at risk.

Taken together, these findings indicate that income, rather than age, is the main driver of heterogeneity in the price response to flood risk. While a full interpretation is beyond the scope of this paper, the observed patterns may reflect differences in financial constraints, information acquisition, education, or bargaining power across income groups.

\subsection{Low-income buyers are more likely to move to at-risk areas}

We now examine whether households purchasing homes in flood-risk areas differ systematically in income from those purchasing properties elsewhere, addressing our final research question. A simple comparison of group means shows that households buying in at-risk areas have an average monthly income of 2,898\euro, compared to 2,953\euro for households buying in non-risk areas—a difference of about 1.9\%. However, this unconditional comparison may be misleading, as flood-risk exposure is not randomly distributed across space and is more prevalent in relatively affluent regions of Italy, such as the Po Valley.

Table~\ref{tab:Income_baselineEMDAT} reports estimates of the effect of flood risk on buyers’ income using Equation~\eqref{eq:Incomebaseline}. Focusing on the coefficient on flood risk, Model~1 shows that, on average across Italy, buyers purchasing properties in flood-risk areas have incomes approximately 0.9\% lower than those purchasing in non-risk areas. Model~2 reveals substantial heterogeneity across awareness levels: income differences are larger and statistically significant in high- and medium-awareness contexts (–1.4\% and –1.2\%, respectively), while the effect becomes small and statistically insignificant under low-awareness conditions.

\begin{figure}
    \centering
    \includegraphics[width=0.75\linewidth]{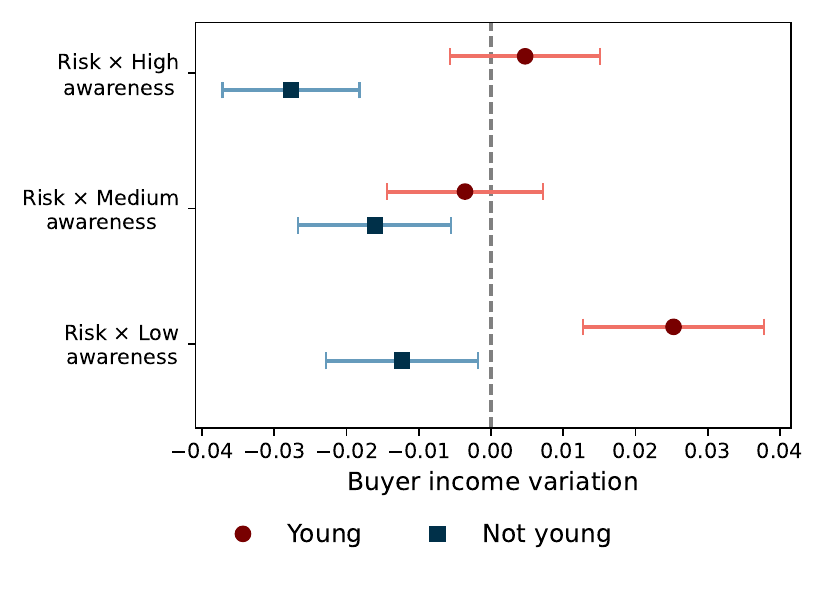}
    \caption{Flood risk effect on buyers' income disaggregated across awareness and mortgage applicant age, Equation~\eqref{eq:incometriple}.}
    \label{fig:income_triple}
\end{figure}

We further explore heterogeneity by buyers’ age by estimating Equation~\eqref{eq:incometriple}. The results, shown in Figure~\ref{fig:income_triple}, indicate pronounced age-related differences. Among \texttt{NotYoung} buyers, incomes of those purchasing homes in flood-risk areas are approximately 3\% lower than the incomes of buyers that do not purchase properties at risk in high-awareness settings. While this income gap narrows as awareness declines, it remains negative across all awareness levels. In contrast, results for \texttt{Young} buyers are less consistent and in several cases not statistically significant.

Overall, these findings suggest that income-based sorting into flood-risk areas is particularly pronounced among older households. This pattern is likely shaped by differences in financial constraints, information, and bargaining power across demographic groups, consistent with the heterogeneity found in Section~\ref{sec:hedonic_risk}.

\begin{table}[]
\caption{Flood risk effect on buyers' incomes, Equation~\eqref{eq:Incomebaseline}. Models 1-2 report the effect of the risk on buyers' income at the national scale (Model 1) and disaggregated by the level of flood awareness at the time of the purchase (Model 2, with half life time equal to 10 years). *** $p < 0.01$, ** $p < 0.05$, * $p < 0.1$.}
\centering
\begin{tabular}{l|cc}
\hline
                                                                                   & \multicolumn{2}{c}{Dep. var.: log(\textit{I})}                                                                     \\ \hline
\textbf{Model id}                                                                  & 1                                                                & 2                                                                \\
\textbf{Risk}                                                                      & \begin{tabular}[c]{@{}c@{}}-0.009$^{***}$\\ (0.003)\end{tabular} &                                                                  \\
\textbf{\begin{tabular}[c]{@{}l@{}}Risk $\times$ High \\ awareness\end{tabular}}   &                                                                  & \begin{tabular}[c]{@{}c@{}}-0.014$^{***}$\\ (0.004)\end{tabular} \\
\textbf{\begin{tabular}[c]{@{}l@{}}Risk $\times$ Medium \\ awareness\end{tabular}} &                                                                  & \begin{tabular}[c]{@{}c@{}}-0.012$^{***}$\\ (0.005)\end{tabular}  \\
\textbf{\begin{tabular}[c]{@{}l@{}}Risk $\times$ Low \\ awareness\end{tabular}}    &                                                                  & \begin{tabular}[c]{@{}c@{}}0.002$^{}$\\ (0.005)\end{tabular}     \\ \hline
\textbf{Observations}                                                              & \multicolumn{2}{c}{552728}                                                                                                          \\
\textbf{R\textsuperscript{2}}                                     & \multicolumn{2}{c}{0.287}                                                                                                           \\ \hline
\textbf{Spatial fix. effects}                                                      & \multicolumn{2}{c}{OMI zone}                                                                                                        \\
\textbf{N. clusters}                                                               & \multicolumn{2}{c}{20297}                                                                                                           \\
\textbf{Clustered s.e.}                                                            & \multicolumn{2}{c}{\checkmark}                                                                                       \\ \hline
\end{tabular}
\label{tab:Income_baselineEMDAT}
\end{table}

\FloatBarrier

\section{Conclusion}
\label{sec:conclusion}
This is the first work that systematically analyzes the impact of flood risk on home prices throughout Italy. Our study reveals three key findings. First, unlike much of the existing literature, we find that even the largest recent flood in Italy--the 2023 Emilia-Romagna flood--did not lead to additional price declines for at-risk homes that were not directly hit. Second, while homes in flood-prone areas generally sell at a small discount of about 1\%, this penalty is concentrated in regions with high awareness conditions (i.e., frequently flooded), where at-risk homes sell at up to 4\% less than safe homes. Third, only high-income buyers obtain a discount when purchasing at-risk properties, while lower-income buyers are more likely to settle in flood-prone areas, suggesting a socioeconomic sorting effect.

To make sense of these findings, we hypothesize that it is the repeated exposure to floods, rather than isolated events, what ultimately leads to the capitalization of flood risk in home prices. We argue that, with no institutional requirements for flood risk disclosure or insurance that encourages the sharing of information regarding flood risk, households rely on the \textit{historical memory} of major floods to identify the most at-risk areas. Finally, we find that lower-income buyers are more likely to settle in flood-prone areas, highlighting potential socioeconomic disparities in climate risk exposure, especially for the next years. 

Thus, our paper contributes to kick off the discussion on how to enhance public awareness on home flood risk, prompting the government and private sector to design effective policies. These could include making flood risk disclosure mandatory, subsidizing flood insurance programs, and offering incentives for homeowners to invest in flood resilience measures. Future research should address how effective these policies are in making buyers consider flood risk in their purchasing decisions and mitigating the clustering of lower-income households in flood-prone areas.

{\footnotesize
\bibliographystyle{econ}
\bibliography{refs}
}

\FloatBarrier
\clearpage
\appendix
\setcounter{section}{0}
\setcounter{figure}{0}
\setcounter{table}{0}

\makeatletter
\renewcommand{\thesection}{\Alph{section}}
\renewcommand \thetable{A\arabic{table}}
\renewcommand \thefigure{A\arabic{figure}}

\section{Intesa Sanpaolo data}
\label{sec:ispdata}

\subsection{A more detailed description}
\label{sec:datadescription}
Here we describe in more detail the variables available in each of the two ISP datasets provided.\footnote{We only describe the variables used in this paper, however, these datasets contain several other variables such as information on the conditions of each mortgage.} We first give an intuitive description of the content of the variable, then indicate its type and finally, when the variable is categorical or binary, list its main categories.

In the \texttt{Contracts} dataset the id variable is the unique mortgage contract code. The other relevant variables are:
\begin{itemize}
\item Type of mortgage applicant; categorical; with levels: single physical person, joint application by two physical persons, juridical person. 
\item Application status; binary; whether the mortgage has been issued, or the application is still under review.
\item Application type; binary; whether the application is for a mortgage or for a loan for building a home with the goal of reselling it to final consumers.
\item Auction flag; binary; whether the housing unit being given as collateral is sold as a foreclosure auction.
\item Goal of the mortgage; categorical; with levels: purchase of 1st or 2nd home, subrogation of a previous mortgage from a bank outside the ISP group, renovation, etc. 
\item Issuance date; date.
\item \emph{Mutuo giovani}; binary, whether the mortgage has incentives for young applicants. 
\item Date of construction of the building; numeric (year). 
\item Transaction price; numeric (euros).
\item Applicant income; numeric (euros). It is the income of the applicant, or the sum of the incomes of the applicants when the application is joint.
\item Latitude and longitude of the main housing unit given as collateral; numeric. 
\end{itemize}

In the \texttt{Cadaster} dataset the id variable is the unique cadastral unit. The other relevant variables are:
\begin{itemize}
\item Floor area; numeric (m$^2$)
\item Cadastral category; categorical. The cadastral code is given by the Italian tax agency to measure the taxable value of housing units and distinguishes between apartment, detached housing unit, garage, basement, etc.\footnote{The meaning of some of the residential cadastral units, which are the ones most relevant for our study, is the following. A01: Luxury residences; A02: Standard residences; A03: Low-cost residences; A04: Working-class residences; A05: Substandard residences; A06: Rural residences; A07: Small villas; A08: Villas; A09: Castles, artistic and historical palaces; A10: Offices or private practices; A11: Typical local residences.} 
\item Energy class, following the 2015 European Union Energy Performance of Buildings Directive; categorical; with levels: from A4 (the highest) to G (the lowest).
\item Air conditioned area: numeric (m$^2$).
\item Floor: categorical.
\end{itemize}

We processed and cleaned these data, to create a single database where the unit is the mortgage contract and the housing unit given as collateral combines information from all linked cadastral units. In the following, we list a few steps:
\begin{enumerate}
\item We built a purchase flag that takes value one when the goal of the mortgage is to purchase a housing unit (as opposed to, e.g., subrogation or renovation). Then we discarded the mortgages not related to purchases.
\item We discarded the mortgages with missing coordinates, or coordinates located outside Italy, that were given to juridical persons, that are still under review and that were issued to construct housing units with the goal of reselling them.
\item We set as missing values the values of floor area and transaction price that are zero. 
\item We summed the floor areas of the residential cadastral units that were given as collateral for the mortgage. This excludes, for instance, the basement or garage of a building, but it sums the floor area of, for example, two apartments that were independent cadastral units at the time of purchase but then are joined following renovation work. 
\item We created a garage flag that takes value one if at least one cadastral unit given as collateral is C06, i.e. a garage or a private parking space.
\item We created an additional annex flag that takes value one if at least one cadastral unit given as collateral is C02, i.e. a basement, an attic, or similar.
\item We associated to each mortgage contract the energy class of the cadastral unit with the highest value from the \texttt{Cadaster} dataset.
\item We built an air conditioning flag that takes value one when the air-conditioned area is larger than zero.
\item We took the raw data on the floor, selected the floor of the cadastral unit with the highest value, and manually made the values of the floor consistent with one another.\footnote{For example we set ``T'', ``TERRA'', ``Terra'', ``t'', ``PIANO TERRA'', ``rialzato'', ``RIALZATO'', ``PIANO RIALZATO'', ``rialzat'', ``T-S1'' all equal to the ground floor.} Since some housing units are divided among multiple floors, we created a flag that takes the value one if it is on multiple floors, and a numeric variable for the floor that takes the value of the minimum floor in the case of multi-floor apartments.
\end{enumerate}

\begin{longtable}{lr}
\caption{Summary statistics of categorical variables. We show the top-10 most frequent categories. The share across all samples is reported among brackets.} \label{tab:summary_statistics_categorical} \\
\toprule
\midrule
\endfirsthead
\caption[]{Summary statistics of categorical variables. We show the top-10 most frequent categories. The share across all samples is reported among brackets.} \\
\toprule
\midrule
\endhead
\midrule
\multicolumn{2}{r}{Continued on next page} \\
\midrule
\endfoot
\bottomrule
\endlastfoot
\textbf{Municipality of Property} &  \\
Roma & 32830 (5.94\%)\\
Milano & 24206 (4.38\%)\\
Torino & 12658 (2.29\%)\\
Genova & 9716 (1.76\%)\\
Napoli & 7222 (1.31\%)\\
Firenze & 7091 (1.28\%)\\
Bologna & 5806 (1.05\%)\\
Bari & 5663 (1.02\%)\\
Palermo & 3777 (0.68\%)\\
Venezia & 3586 (0.65\%)\\
Missing & 0 (0\%)\\
\textbf{Cadastral code} &  \\
A02: standard residences & 267974 (48.48\%)\\
A03: low-cost residences & 202693 (36.67\%)\\
A07: small villas & 43614 (7.89\%)\\
A04: working-class residences & 37118 (6.71\%)\\
A05: substandard residences & 779 (0.14\%)\\
A01: luxury residences & 271 (0.05\%)\\
A08: villas & 125 (0.02\%)\\
A10: offices or private practices & 97 (0.018\%)\\
A06: rural residences & 76 (0.013\%)\\
A11: typical local residences & 12 (0.002\%)\\
Missing & 0 (0\%)\\
\textbf{Floor (categorical)} &  \\
1 & 28423 (5.14\%)\\
0 & 24600 (4.45\%)\\
2 & 18175 (3.29\%)\\
3 & 10726 (1.94\%)\\
4 & 6087 (1.10\%)\\
0-1 & 5905 (1.07\%)\\
5 & 3635 (0.66\%)\\
6 & 2133 (0.39\%)\\
0-1-2 & 1345 (0.24\%)\\
Missing & 447545 (80.97\%)\\
\textbf{Flag multiple floors} &  \\
No & 96069 (17.38\%)\\
Yes & 9147 (1.65\%)\\
Missing & 447545 (80.97\%)\\
\textbf{Garage Flag (Indicates if the property includes a garage)} &  \\
Yes & 282883 (51.18\%)\\
No & 269878 (48.82\%)\\
Missing & 0 (0\%)\\
\textbf{Additional Basement or Attic Flag (annex)} &  \\
No & 473369 (85.64\%)\\
Yes & 79392 (14.36\%)\\
Missing & 0 (0\%)\\
\textbf{Energy Efficiency Class} &  \\
G & 142384 (25.76\%)\\
F & 69803 (12.63\%)\\
E & 52573 (9.51\%)\\
D & 37175 (6.73\%)\\
A4 & 21009 (3.80\%)\\
C & 19990 (3.62\%)\\
B & 17037 (3.08\%)\\
A2 & 12844 (2.32\%)\\
A1 & 8736 (1.58\%)\\
Missing & 162907 (29.47\%)\\
\textbf{Air Conditioning Flag (Indicates if the property has air conditioning)} &  \\
No & 177367 (32.09\%)\\
Yes & 41903 (7.58\%)\\
Missing & 333491 (60.33\%)\\
\textbf{Flood risk} &  \\
No risk & 425753 (77\%) \\
Low risk & 63908 (11\%) \\
Medium risk & 42994 (8\%) \\
High risk & 20106 (4\%) \\
Missing & 0 (0\%) \\
\textbf{Young Buyers' Mortgage (Mutuo Giovani)} &  \\
No & 329705 (59.65\%) \\
Yes & 223023 (40.35\%) \\
Missing & 0 (0\%) \\
\end{longtable}

\subsection{Are the data representative of the Italian housing market?}
\label{sec:representativity}

We want to establish if the sample of mortgages issued by ISP is representative of the Italian housing market. We compare the ISP dataset to multiple public data sources at various spatial and temporal aggregation scales, focusing on the number of transactions, prices and incomes. Our main data sources for comparison are from \emph{Osservatorio del Mercato Immobiliare} (OMI), the housing market observatory of the Italian Tax Office, and the \emph{Ministero dell'Economia e delle Finanze} (MEF), the ministry of economics and finance. OMI provides information on the number of transactions, disaggregated by municipality and home size. OMI also provides information on prices, disaggregated by cadastral category (apartment, independent house, etc.) and OMI microzone, roughly corresponding to a neighborhood. MEF provides, by income class\footnote{Total income is divided into eight income groups: 'less than zero euros', 'from 0 to 10,000 euros', 'from 10,000 to 15,000 euros', 'from 15,000 to 26,000 euros', 'from 26,000 to 55,000 euros', 'from 55,000 to 75,000 euros', 'from 75,000 to 120,000 euros', 'more than 120,000 euros'.}, number of individuals and total income, yearly at municipality level. 

We first check if the number of transactions in ISP data aligns with OMI. Note that ISP data report information on the transactions connected to a mortgage, so they only represent the transactions of ISP clients who obtained a loan. By contrast, OMI data are retrieved from notary deeds and so describe the universe of transactions in Italy.\footnote{OMI provides the normalized number of transactions (NTNs), i.e., the number of transactions scaled by the fraction of the property that has been transacted at the municipality scale, also considering the home size (square meters, m$^2$).} At the national scale, our data make up 10.7\% of transactions across the Italian housing market; this percentage rises to 14.4\% when we evaluate only the mortgage-financed transactions.\footnote{Between 2016 and 2022 (the last year for which we have OMI data), our dataset records 465,037 mortgages given to purchase a housing unit in Italy. In the same period, OMI declares 4,336,331 transactions, giving a 10.7\% share. The share of mortgage-financed transactions comes from the Bank of Italy survey on the housing market (\url{https://www.bancaditalia.it/pubblicazioni/sondaggio-abitazioni/index.html}). Taking a simple mean across all quarterly surveys between 2016 and 2023, the share of transactions financed by a mortgage is 74.5\%, leading to our 14.4\% estimate on the share of ISP mortgages.}

We also want to identify potential biases in the size distribution of the homes bought with an ISP mortgage with respect to the whole distribution in the Italian market. The OMI dataset reports the size of the housing units using five different size ranges: $0<$m$^2<50$, $50<$m$^2<85$, $85<$m$^2<115$, $115<$m$^2<145$ and m$^2>145$. To compare both datasets, we aggregate ISP data to the OMI home size ranges and introduce the variable $r(c)$ that describes the share of transactions in each home size range. Mathematically speaking, $r(c)$ is
\begin{equation}
r(c) = \frac{n(c)}{\sum_{c'}n(c')},
\label{eqn:sharebym2}
\end{equation}
where $n(c)$ is the number of transactions in the home size range $c$.
Using Equation~\ref{eqn:sharebym2}, we compute $r(c)$ for both ISP and OMI data. 

\begin{figure}[!h]
    \centering
    \includegraphics[width=0.8\textwidth]{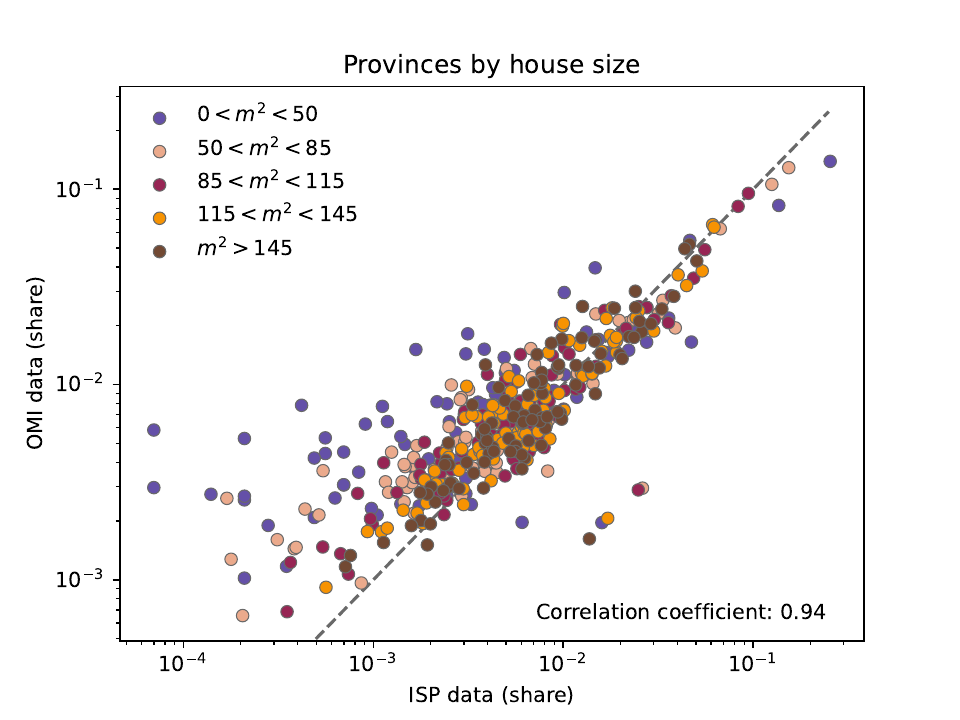}
    \caption{Comparison of transaction shares between ISP and OMI data for provinces and home size ranges. Each dot represents a province considering a specific home size range. The dashed gray line represents the identity line, where ISP and OMI datasets present equal shares. Both axes are on a logarithmic scale.}
    \label{fig:scatter_m2}
\end{figure}

Figure~\ref{fig:scatter_m2} shows the transaction shares grouped by provinces and home size ranges.\footnote{We computed Equation~\eqref{eqn:sharebym2} for all provinces $p$ and size ranges $c$ (i.e., $r(p,c) = n(p,c)/\sum_{c'} n(p,c')$) for the ISP and OMI datasets.} We observe a correlation coefficient of 0.94 (i.e., strong correlation) between the datasets. We also note that smaller housing units tend to have a larger deviation from the identity line, in the sense that ISP data seem under-represented. When looking at larger housing units, the agreement of the OMI and ISP data increases (the points align on the identity line). This is reasonable, as it is more likely to apply for a mortgage to afford large homes.

We now check if the prices computed from ISP data are similar to the prices that are given in the administrative data by OMI. For a combination of OMI microzone, maintenance status and cadastral category, OMI provides an estimate of the maximum and minimum price per m$^2$. We consider the midpoint of this interval as the mean home price, and focus on the normal maintenance status, which makes up 88\% of the data. We compute the mean price over the 2016-2023 period, for each combination of OMI microzone and cadastral category. The prices are highly correlated in the two datasets (Pearson correlation coefficient: 0.77), although the correlation is not perfect. 

We are also interested in comparing prices at the municipality level. The only way to know the OMI mean price per m$^2$ in a municipality is to compute the weighted mean of the price per m$^2$ in each OMI microzone in the municipality, where weights are the fraction of transactions in the OMI microzone compared to the municipality.\footnote{Since OMI does not provide data on transactions at a level below the municipality, we use instead the fraction of housing units in each OMI microzone as a proxy. In turn, this is computed from census data by aggregating all the census tracts that correspond to any given OMI microzone.} To increase the sample size, we pool across cadastral categories and years. 

\begin{figure}[!h]
    \centering
    \includegraphics[width=\textwidth]{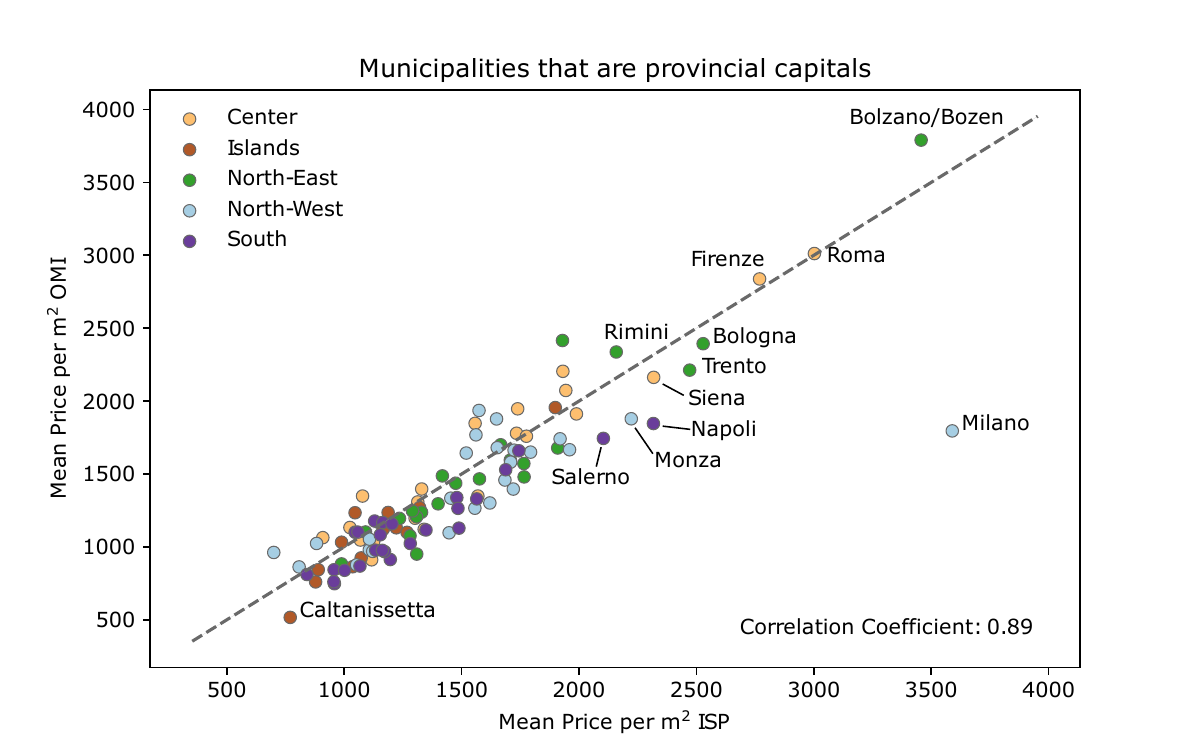}
    \caption{Comparison of prices per m$^2$ between ISP and OMI data. Each dot corresponds to a municipality that is a provincial capital and shows the mean price per m$^2$ over the 2016-2023 period. It is colored according to the macro-area of Italy the municipality belongs to. The dashed black line shows the identity line, where ISP and OMI price data are identical. We label a few municipalities and report the correlation coefficient between ISP and OMI data.}
    \label{fig:representativeness_prices_comuni_scatter}
\end{figure}

Figure \ref{fig:representativeness_prices_comuni_scatter} shows the comparison of the municipality- and time-averaged price per m$^2$ in ISP and OMI data. Compared to the result at the OMI microzones level, the correlation is much higher, suggesting that ISP data aggregated at the municipality level are highly representative of the Italian housing market. In fact, this figure suggests that ISP data may also be more informative, as the case of Milano suggests. For Milano, ISP data indicate a mean price of over 3500\euro per m$^2$, while OMI data indicate a price of slightly more than 1500\euro per m$^2$. Given that Milano is one of the most expensive cities in Italy, the ISP data are likely to be more accurate.

\begin{figure}[!h]
	\centering
	\includegraphics[width=\textwidth]{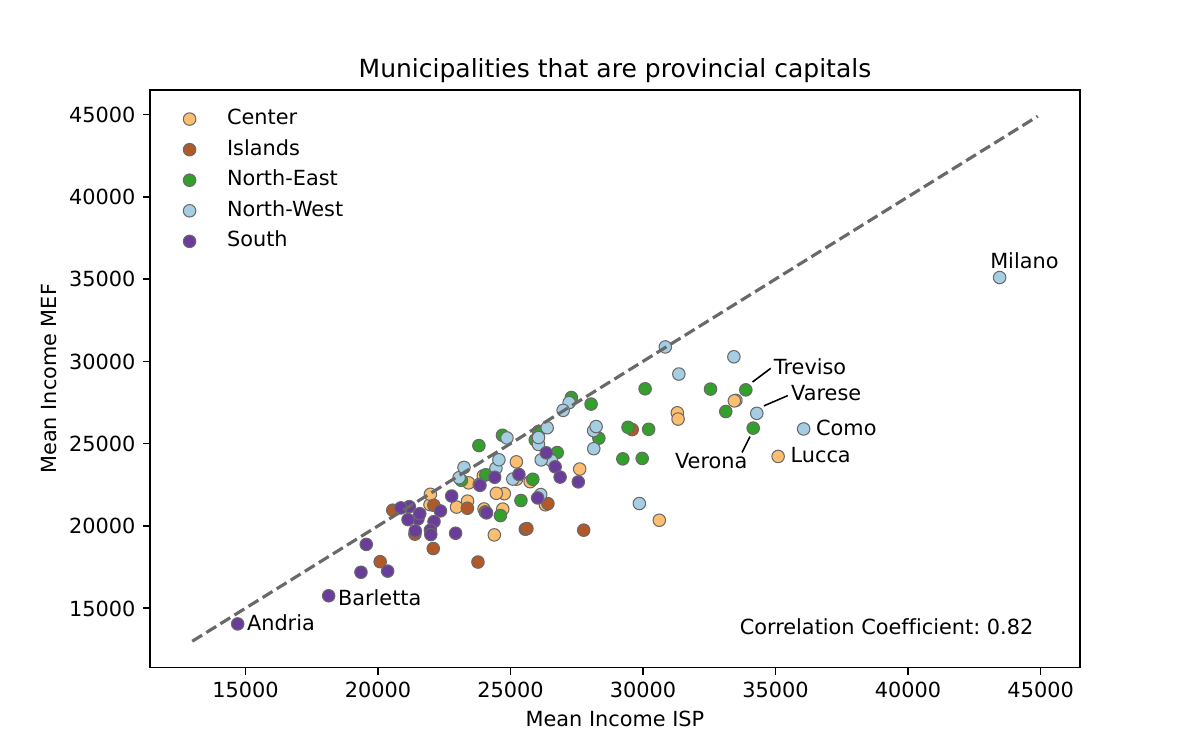}
	\caption{Comparison of incomes between ISP and MEF data. Each dot corresponds to a municipality that is a provincial capital and shows the mean income over the 2016-2021 period. It is colored according to the macro-area of Italy the municipality belongs to. The dashed black line shows the identity line, where ISP and MEF income data are identical. We label a few municipalities and report the correlation coefficient between ISP and MEF data.}
\label{fig:representativeness_incomes_comuni_scatter_provcap}
\end{figure}

We finally check how the incomes declared by ISP mortgage applicants compare to data on incomes made publicly available by MEF. We expect two substantial biases that may offset each other. On one hand, the population of ISP mortgage applicants is likely to have a higher income than the general population. On the other hand, ISP incomes are net while MEF incomes are gross.\footnote{Moreover, MEF incomes always refer to individuals, whereas ISP incomes refer to individuals when the applicant is a single physical person and to households in case of joint applications. We address this issue by dividing applicant income by two in case of joint applications, as the vast majority of joint applications are filed by two physical persons. 
} 
Focusing on the provincial capitals, Figure \ref{fig:representativeness_incomes_comuni_scatter_provcap} shows that there is a very strong correlation between ISP and MEF incomes (Pearson correlation: 0.82). Moreover, ISP incomes are slightly above MEF incomes, suggesting that the overestimation of incomes because ISP focuses on mortgage applicants rather than on the general population is slightly stronger than the underestimation of incomes due to ISP considering net rather than gross incomes. 

\section{Supplementary results}

\subsection{Specific flood events}\label{supp:ER_extra}

We evaluated the robustness of our findings on the insignificant effect of specific flood events in three ways: (i) we considered other regression specifications; (ii) we only focused on ground-floor apartments or independent houses, which are more likely to be damaged from floods than homes on upper floors; and (iii) we explore the changes in home prices after the second and third largest flood events in the studied period.

\subsubsection*{Testing different regression specifications}
We evaluate the robustness of our results by considering alternative specifications for the variables describing: home size (square meters, m$^2$), floor, and spatial fixed effects. Specifically, we consider three specifications for home size: the m$^2$ as reported in our data (indicated as linear), the logarithm of the home size (log), and ranges of home size (bins), coherently with the \emph{Osservatorio del Mercato Immobiliare} (OMI), grouping homes by size (m$^2$$<$50, 50$<$m$^2<$85, 85$<$m$^2<$115, 115$<$m$^2<$145, and m$^2>$145). For the floor, our specifications include the floor as reported in the ISP data and a binned version, where all the homes with floor numbers larger or equal to 4 have been grouped. We consider also three possible spatial fixed effects in our regression, namely the census tract, the OMI microzone, and the municipality (Mun.).  Moreover, we also evaluate the impact of including or removing data outliers in the regression outcomes. Specifically, when we remove the outliers from the data, we do not consider the top and bottom 0.1\% of the data regarding the sale price, home size and and household's income. Therefore, we investigate 36 different parameters configurations resulting from the possible specifications of home size, floor, spatial fixed effect, and outliers removal. The configuration reported in the main text considers the logarithm of home size, binned floor, fixed effects at OMI microzone, and outliers have been removed.

\begin{figure}
    \centering
    \includegraphics[width=1\linewidth]{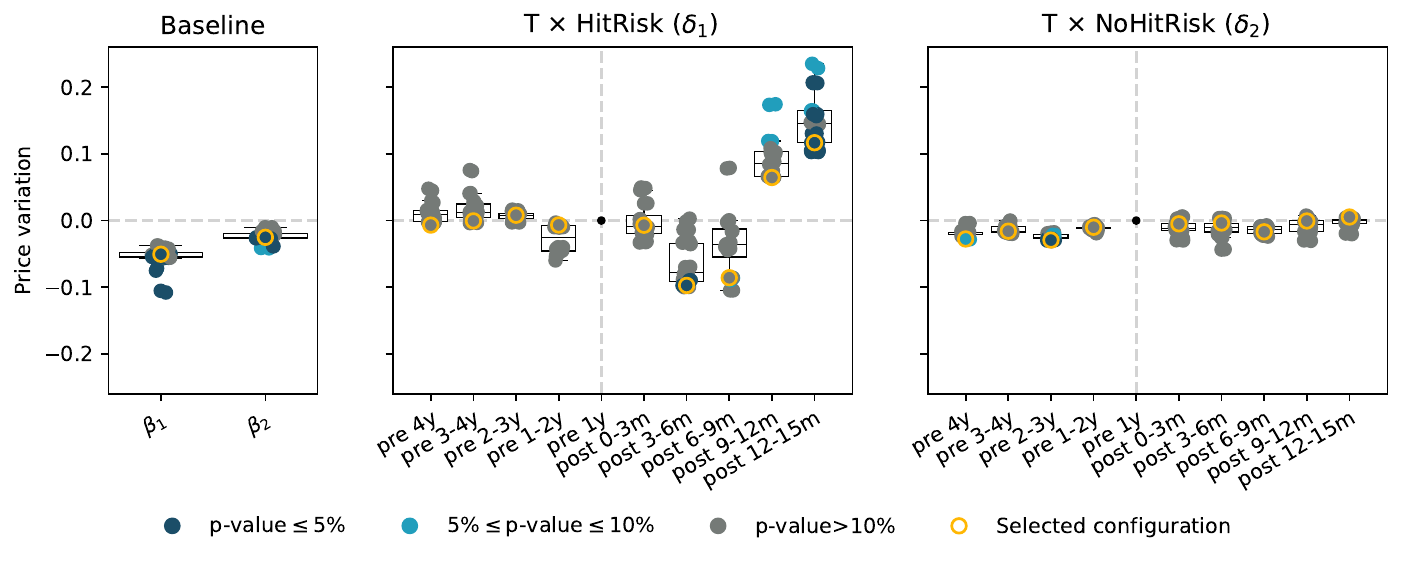}
    \caption{Home price variation across specifications using the Difference in Difference analyses (Equation~\ref{eq:diffindiff}) for the Emilia-Romagna flood. Each dot represents a parameter configuration, the color reports the statistical significance. The dots with the yellow edges identify the parameter configuration presented in the main text. R$^2$ [min, max]~=~[0.55, 0.86]; observations [min, max]~=~[35623, 35815].}
    \label{fig:EmiliaRomagnaFlood_quarter}
\end{figure}

Figure~\ref{fig:EmiliaRomagnaFlood_quarter} reports the robustness analysis for the 2023 Emilia-Romagna flood (Equation~\ref{eq:diffindiff}). In this case, we tested only OMI microzone and Census tract as spatial fixed effects, resulting in 24 parameter configurations.\footnote{We omitted the municipality fixed effects because they are already considered in the definition of the ``hit" municipalities and so including the fixed effects would lead to colinearity.} The results for the selected configuration agree with those for the other specifications, supporting the results reported in the main text. 

\subsubsection*{Zoom in on ground-floor and independent homes}
\begin{figure}[ht]
    \centering
    \includegraphics[width=\linewidth]{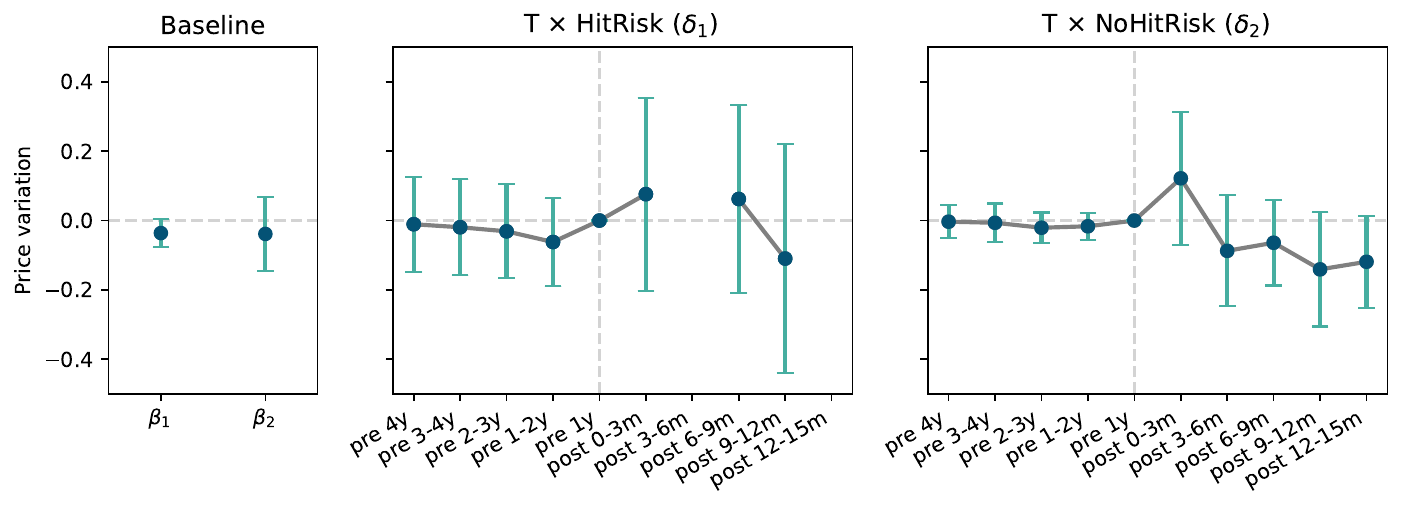}
    \caption{Time evolution of the home price variation for the 2023 Emilia-Romagna flood, considering only transactions related to ground-floor apartments or independent houses. Referring to Equation~\eqref{eq:diffindiff}, the left panel reports the coefficients $\beta_1$ and $\beta_2$, the central and right panels show the interaction coefficients $\delta_1$ and $\delta_2$ up to 15 months after the flood of Emilia-Romagna 2023. Each point represents the average value of the regression coefficient and the error bars indicate the 95\% confidence intervals, considering the spatial fixed effect at the OMI microzone level. Number of \texttt{HitRisk} homes sold before the 2023 Emilia-Romagna flood: 451; 10 homes were sold after the event. For the \texttt{NoHitRisk} homes, 1999 homes were sold before the event and 80 after the flood (the total number of homes at ground-floor or villas in the affected municipalities is 3276). We had no data for the \textit{post 3-6m} bin for directly hit homes. R$^2$ = 0.74, observations = 8634.}
    \label{fig:vilette_dif}
\end{figure}

Figure~\ref{fig:vilette_dif} presents the results of estimating Equation~\eqref{eq:diffindiff} only on homes located on the ground floor or independent houses in the Emilia-Romagna region. In this case, our results lack statistical significance, probably due to the limited number of transactions for these housing units (less than a quarter of the total number of transactions in the Emilia-Romagna region). Despite this limit, we note that ground floor homes and independent houses in the \texttt{NoHitRisk} class and sold after the flood do not present a significant price variation, supporting our claim that individual flood events do not cause changes in the price dynamics for non-affected homes.

\subsubsection*{Other floods events}
Aiming to extend our results on the 2023 Emilia-Romagna flood to the entire Italian housing market, we repeated the analysis considering the 2016 Piemonte-Liguria (code EMSR192) and 2023 Toscana (EMSR705) floods. As reported in Table~\ref{tab:flood_events}, the considered events are the only two other floods --according to Copernicus-- that had a substantial impact on the homes in our dataset. The Piemonte-Liguria and Toscana floods involve only few homes that were hit and sold after the events (28 and 4, respectively), suggesting that estimates would be very noisy. On top of that, we are particularly interested in the impact on homes that were not directly hit, for which we have a set of transactions comparable to the Emilia-Romagna flood (EMSN154). In fact, 3968 \texttt{NoHitRisk} homes were sold after the Piemonte-Liguria flood and 1810 from the flood in Toscana in 2023 (for the 2023 Emilia-Romagna event, EMSN154, we have 1247 transaction for the \texttt{NoHitRisk} homes). 

\begin{figure}[ht]
    \centering
    \includegraphics[width=1\linewidth]{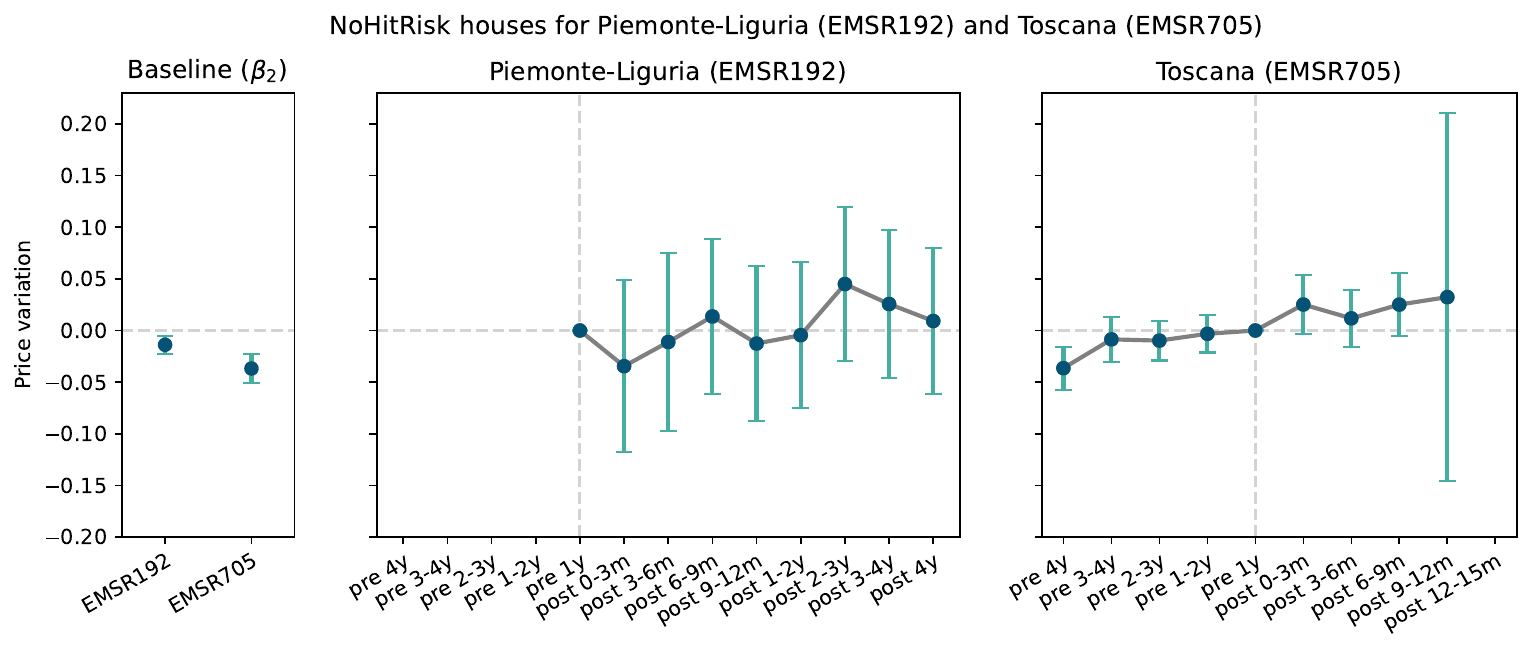}
    \caption{Time evolution of the home price variation for the 2016 Piemonte-Liguria (EMSR192) and 2023 Toscana (EMSR705) flood events. Referring to Equation~\eqref{eq:diffindiff}, the left panel reports the coefficients $\beta_2$ for both events, the central and right panel show the interaction coefficients $\delta_2$ for the Piemonte-Liguria and Toscana events, respectively. Each point represents the average value of the regression coefficient and the error-bars indicate the 95\% confidence intervals, considering fixed effect at the OMI microzone. Since our dataset covers mortgage-covered transaction between 2016 and August 2024, no data points are available before one year from the Piemonte-Liguria events and after one year from the Toscana flood. For Piemonte-Liguria (EMSR192): R$^2$ = 0.77 observations~=~231147. For Toscana (EMSR705): R$^2$=0.71, observations~=~46624.}
    \label{fig:other_floods}
\end{figure}

Figure~\ref{fig:other_floods} shows the results for the \texttt{NoHitRisk} homes. Coherently with our baseline estimates and the 2023 Emilia-Romagna flood, there is a statistically significant ``baseline'' reduction in prices for homes located in risk areas, as indicated by the $\beta_2$ coefficients shown in the left panel. Regarding \texttt{NoHitRisk} homes, no further price variation is observed after the considered flood events, further supporting our argument on the need of repeated events for observing home price reductions.\footnote{Table~\ref{tab:statistic_test_pre_post} compares the statistical properties of the considered homes one year before and one year after the Piemonte-Liguria and Toscana flood events.}

\begin{table}[t]
    \centering
    \caption{Characteristics of the major flood events reported by Copernicus (2016-2023). The column ``\# homes aff. mun.'' represents the number of homes in municipalities affected by each flood. The column ``\# homes at risk" counts the homes in the affected municipalities that are at flood risk according to the ISPRA maps. The column ``\# hit homes'' indicates the number of homes within the ISP dataset located inside the maximum flooded area for each event.}
    \begin{tabular}{lllrrrr}
    \toprule
    Code & Location & Date & \makecell{\# homes \\in aff. mun.} & \makecell{\# homes at\\risk} & \makecell{\# hit \\homes} \\
    \midrule
    EMSR192 & Piemonte \& Liguria & 24th Nov. 2016 & 25406 & 4077 & 28 \\ 
    EMSR260 & North-West area & 12th Dec. 2017 & 5825 & 4607 & $\leq4$ \\ 
    EMSR359 & Emilia-Romagna & 15th May 2019 & 5514 & 4185 & $\leq4$ \\ 
    EMSR483 & Calabria & 22nd Nov. 2020 & 379 & 94 & $\leq4$ \\ 
    EMSR486 & Sardegna & 28th Nov. 2020 & 121 & 8 & $\leq4$\\ 
    EMSR487 & Emilia-Romagna & 6th Dec. 2020 & 2161 & 1833 & $\leq4$ \\ 
    EMSR488 & Lazio & 9th Dec. 2020 & 149 & 72 & $\leq4$ \\ 
    EMSR649 & Sicilia & 9th Feb. 2023 & 2534 & 159 & $\leq4$\\
    EMSN154 & Emilia-Romagna & 16th May 2023 & 14977 & 10585 & 1464\\ 
    EMSR705 & Toscana & 2nd Nov. 2023 & 21900 & 18964 & 94\\
    \bottomrule
    \end{tabular}
\label{tab:flood_events}
\end{table}

\begin{table}[]
\caption{Statistical tests comparing home characteristics one year before and after each event. For continuous variables, we applied the Mann-Whitney U-test~\citep{mann1947test} and a paired t-test~\citep{student1908probable}. We also applied the later test to binary variables, together with a $\chi^2$ test~\citep{chi_square1000criterion}. Categorical variables were only assessed through a $\chi^2$. To ensure the validity of the $\chi^2$ test we removed categories with less than 20 samples. Cells report the p-value of the tests, while stars indicate the usual p-value ranges *** $p < 0.01$, ** $p < 0.05$, * $p < 0.1$.}
\label{tab:statistic_test_pre_post}
\centering
\begin{tabular}{llcll}
\hline
Event & Hedonic characteristics & Mann-Whitney & Paired t-test & $\chi^2$ test \\ \hline
\multirow{11}{*}{EMSN154} & Logarithm of square meters & 0.864 & 0.867 &  \\
 & Cadastral code &  &  & 0.745 \\
 & Construction year &  &  & $0.0^{***}$ \\
 & Energy class &  &  & $0.034^{**}$ \\
 & Floor &  &  & 0.655 \\
 & Flag multi floor &  & $0.0^{***}$ & $0.0^{***}$ \\
 & Flag garage &  & 0.727 & 0.744 \\
 & Flag air conditioning &  & $0.0^{***}$ & $0.0^{***}$ \\
 & Flag annex &  & 0.11 & 0.114 \\
 & Flood risk &  & 0.883 & 0.902 \\\hline
\multirow{11}{*}{EMSR192} & Logarithm of square meters & $0.076^{*}$ & $0.023^{**}$ &  \\
 & Cadastral code &  &  & $0.087^{*}$ \\
 & Construction year &  &  & $0.029^{**}$ \\
 & Energy class &  &  & $0.002^{***}$ \\
 & Floor &  &  & $0.028^{**}$ \\
 & Flag multi floor &  & 0.647 & 0.679 \\
 & Flag garage &  & $0.013^{**}$ & $0.013^{**}$ \\
 & Flag air conditioning &  & $0.074^{*}$ & $0.086^{*}$ \\
 & Flag annex &  & $0.0^{***}$ & $0.0^{***}$ \\
 & Flood risk &  & 0.147 & 0.148 \\\hline
\multirow{11}{*}{EMSR705} & Logarithm of square meters & 0.14 & 0.172 &  \\
 & Cadastral code &  &  & $0.008^{***}$ \\
 & Construction year &  &  & $0.01^{***}$ \\
 & Energy class &  &  & $0.0^{***}$ \\
 & Floor &  &  & 0.163 \\
 & Flag multi floor &  & $0.01^{***}$ & $0.0^{***}$ \\
 & Flag garage &  & 0.705 & 0.72 \\
 & Flag air conditioning &  & $0.0^{***}$ & $0.0^{***}$ \\
 & Flag annex &  & 0.903 & 0.926 \\
 & Flood risk &  & 0.134 & 0.14 \\ \hline
\end{tabular}
\end{table}

\FloatBarrier
\clearpage
\subsection{Hedonic regressions}
\label{sec:BaselineEq-SM}

We performed some complementary analyses to validate the specifications chosen for our main regressions. We present four different sets of supplementary results: (i) spatial correlation analysis to check if spatial econometric approaches are needed; (ii) robustness analysis for the risk zone definition; (iii) robustness analysis for the home characteristics included in the baseline regression, (iv) robustness of the historical memory (awareness) specification.

\subsubsection*{Spatial correlation validation}\label{sec:spat_corr}
\label{sec:spatial_correlation}
Samples in geo-localized datasets are likely to exhibit spatial correlation. This violates the assumption of sample independence made in linear regression models and could lead to erroneous estimates if not addressed properly~\citep{anselin2013spatial}. Our hedonic regressions incorporate geographical fixed effects precisely to correct for this influence. However, we must validate whether the fixed effects are capturing completely the intrinsic spatial correlation of the data.

To do so, we applied a spatial autocorrelation test (Global Moran's I)~\citep{moran1950notes} on the residuals of our baseline regression (Model 3 from Table~\ref{tab:NewBaselinePriceTable_EMDAT_numflood}). If the spatial fixed-effect term ($\alpha_k$) correctly captures all spatial dependence in the data, the residuals of the regression ($\epsilon_{ikt}$) should be spatially independent. Meanwhile, if some spatial autocorrelation still remains, it indicates that the hedonic regression is mispecified and spatial approaches should be used instead. 

The Moran's I test requires inverting the spatial weights matrix ($W$) defining the connectivity pattern between samples. In our case, it is not computationally feasible to invert a spatial weights matrix ($W$) for the complete sample size of the baseline regression ($W_{homes \times homes} =552,761 \times 552,761$). Therefore, we averaged the regression residuals of all homes in each municipality and estimated if this average indicator is spatially auto-correlated. ~\footnote{Coarser aggregation levels like the census tract or OMI microzone still result in $W$ matrices too large to be inverted.} The connectivity pattern between municipalities was defined following a queen contiguity criteria, where two municipalities are linked together if there is any point in intersecting in their borders. The resulting binary contiguity matrix was row-normalized to generate the spatial weights matrix $W_{\texttt{municipalities}}$.

The resulting Global Moran's I for our baseline regression (Table~\ref{tab:moran}) is very close to 0. This indicates that the OMI microzone spatial fixed effects are correctly capturing the spatial dependence in the data and thus, our baseline regression is well specified.

\begin{table}[t]
    \centering
    \caption{Test for spatial autocorrelation on the residuals of our baseline regression (corresponding to Equation~\ref{eq:baseline}). Note that we estimated the Global Moran's I at the municipality level assuming a queen contiguity relationship between municipalities~\citep{anselin2009geoda}. Significance levels: *** = p-value$<0.01$, ** = p-value $<0.05$, * = p-value $<0.1$.}
    \begin{tabular}{rl}
    \toprule
    Regression & Global Moran's I  \\
    \midrule
    Model 3 Table~\ref{tab:NewBaselinePriceTable_EMDAT_numflood} & $-0.0590^{***}$ \\ 
    \bottomrule
    \end{tabular}
    \label{tab:moran}
\end{table}

\begin{figure}[ht]
    \centering
    \includegraphics[width=0.6\linewidth]{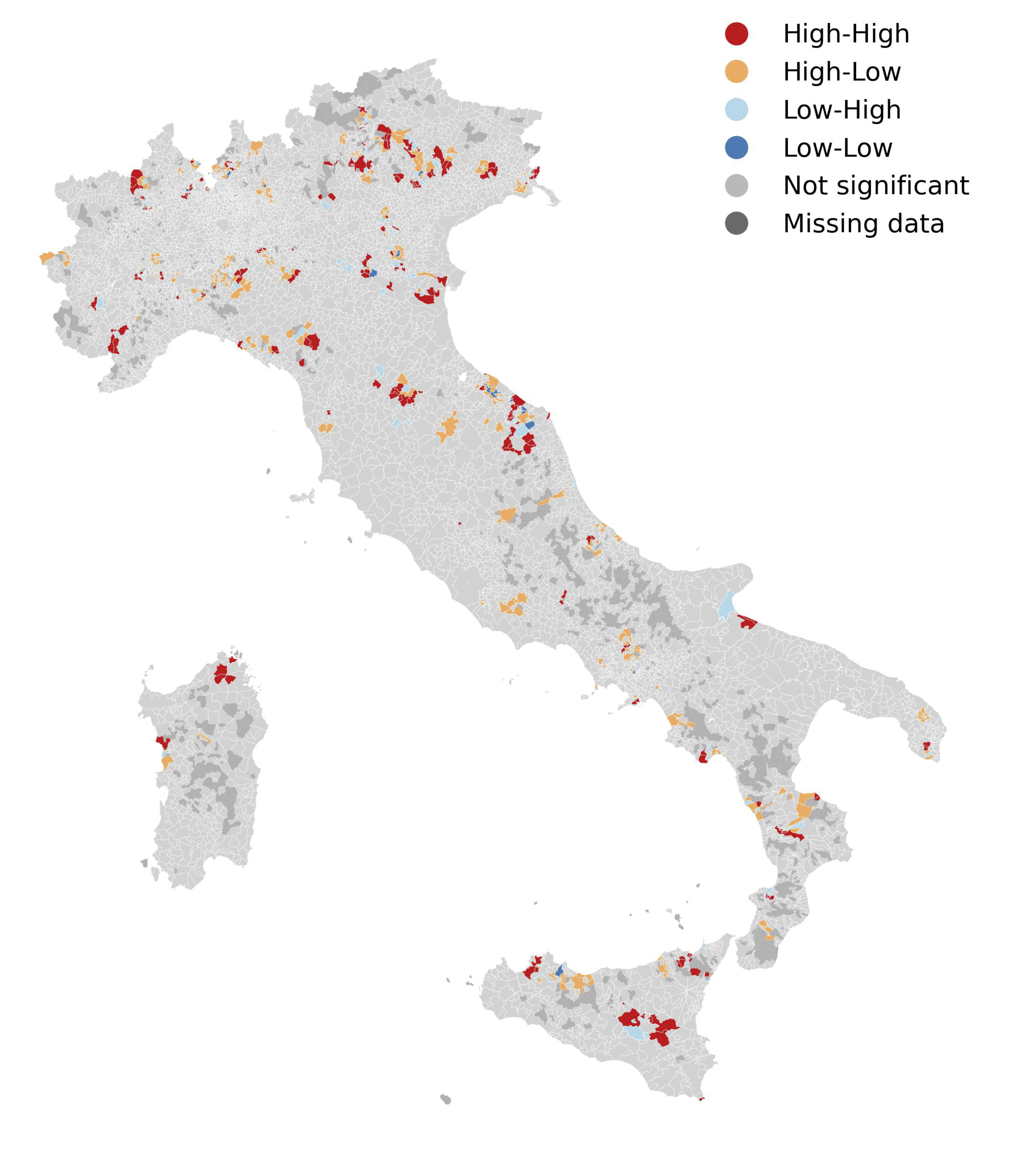}
    \caption{Local Indicators of Spatial Association maps (LISA maps) for the regression residuals for Model 3 from Table.~\ref{tab:NewBaselinePriceTable_EMDAT_numflood}. Municipalities are colored according to their type of local spatial correlation (Local Moran's I). Red (High-High): municipalities with residuals significantly larger than average, surrounded by others alike. Yellow (High-Low): municipalities with larger errors than average, surrounded by municipalities with smaller than average errors. Light blue (Low-High): municipalities with residuals significantly smaller than average, surrounded by others with larger residuals. Dark blue (Low-Low): municipalities with residuals significantly smaller than average, surrounded by others alike. Light gray (Not significant): municipalities with no spatial correlation in the residuals. Dark Gray (Missing data): municipalities without data.}
    \label{fig:moran_risk}
\end{figure}

We also used Local Indicators of Spatial Association maps (LISA maps)~\citep{anselin1995local} to visualize local spatial correlation patterns and validate that there are no specific regions where we systematically have more unreliable results. For all the regressions assessed, we did not observe a significant spatial correlation in most of the municipalities evaluated. Moreover, in the municipalities where the spatial correlation is present, it does not follow a systematic pattern that could indicate unreliable results (see Figure~\ref{fig:moran_risk}). 

\subsubsection*{Validation of the flood risk definition}
For the main analysis we considered an aggregated \texttt{Risk} indicator that combines the three risk levels provided by ISPRA (\texttt{
High}, \texttt{Medium}, \texttt{Low} risk) into a single category. To validate this assumption we performed a robustness analysis where we use the three original categories defined by ISPRA. The equation used is the following: 

\begin{equation}
\log{P_{ikt}} = \alpha_k + \alpha_t + \beta_1 \texttt{HighRisk}_i +  \beta_2 \texttt{MediumRisk}_i + \beta_3 \texttt{LowRisk}_i +\boldsymbol{\gamma}' \textbf{X}_i + \epsilon_{ikt},
\label{eq:risk_levels}
\end{equation}
where \texttt{HighRisk}, \texttt{MediumRisk}, \texttt{LowRisk} reflect dummy variables representing whether a home is located in one of those risk levels according to the ISPRA maps. The rest of the terms are equivalent to the ones in Equation \eqref{eq:baseline}.

\begin{table}[ht]
    \centering
    \caption{Baseline effect of risk on prices disaggregated according to ISPRA risk levels. *** $p < 0.01$, ** $p < 0.05$, * $p < 0.1$.}
    \centering
    
    \begin{tabular}{l|c|c|c}
        \hline
        & \multicolumn{3}{c}{Dep. var.: log(\textit{P})} \\
        \hline
        \textbf{Model id} & 1 & 2 & 3 \\
        \hline
        \textbf{High Risk} & -0.002$^{}$& -0.007$^{}$ & -0.001$^{}$ \\
                             & (0.027)& (0.005) & (0.005) \\[6pt]
                             
        \textbf{Medium Risk} & -0.044$^{**}$& -0.013$^{***}$ & -0.008$^{}$ \\
                             & (0.018)& (0.005) & (0.005) \\[6pt]
                             
        \textbf{Low Risk} & -0.032$^{**}$& -0.009$^{**}$ & -0.012$^{***}$ \\
                             & (0.016)& (0.004) & (0.004) \\[6pt]
        \hline
        \textbf{Observations} & 552761 & 552761 & 552761\\

        \textbf{R\textsuperscript{2}} & 0.669 & 0.755 & 0.848 \\
        \hline
        \textbf{Spatial fixed effects} & Municipality & OMI zone & Census Tract \\
        \textbf{Number of clusters} & 7203 & 20297 & 160630 \\
        \textbf{Clustered s.e.} & \checkmark & \checkmark & \checkmark \\
        \hline
    \end{tabular}
    
    \label{tab:price_risk_levels}
\end{table}

Table~\ref{tab:price_risk_levels} shows the coefficient for the risk indicators when using different types of fixed effects. With the OMI microzone fixed effects we observe that being in a \texttt{LowRisk} flood zone reduces home prices by 0.9\%, while being in a \texttt{MediumRisk} and \texttt{HighRisk} area reduces prices by 1.3\% and 0.7\%, respectively. While it could be expected that being in medium risk zone leads to a higher discount than being in a low risk zone, it is surprising that the discount in high risk zones is the lowest, and not statistically significant. This result makes sense once we consider that only a small number of transactions occurred in high risk zones (4\% of transactions across the whole dataset). In addition, the definition of high risk zone is the one that varies the most across river basin authorities~\citep{trigila2021dissesto}, suggesting that the measurement error for this variable is substantial. This is confirmed when comparing these results with the coefficient for the combined risk indicator $R_i$ (Model 3 in Table~\ref{tab:NewBaselinePriceTable_EMDAT_numflood}). We find that being in a risky area is associated with an average price reduction of 1.0\%, in between the low and medium risk estimates. This further suggests that these two categories are the most relevant ones in the whole data.

\subsubsection*{Testing different regression specifications}
To ensure our regression results are robust, we also validated the impact of modifying the same regression specifications studied in the Emilia-Romagna flood (Supplementary Section \ref{supp:ER_extra}). This includes, the definition of the home size (logarithmic, linear, binned), home floor (binned or logarithmic), fixed effects (Census Tract, OMI microzone, Municipality) and removing (or not) the outliers. 

\begin{figure}[ht]
    \centering
    \includegraphics[width=0.7\linewidth]{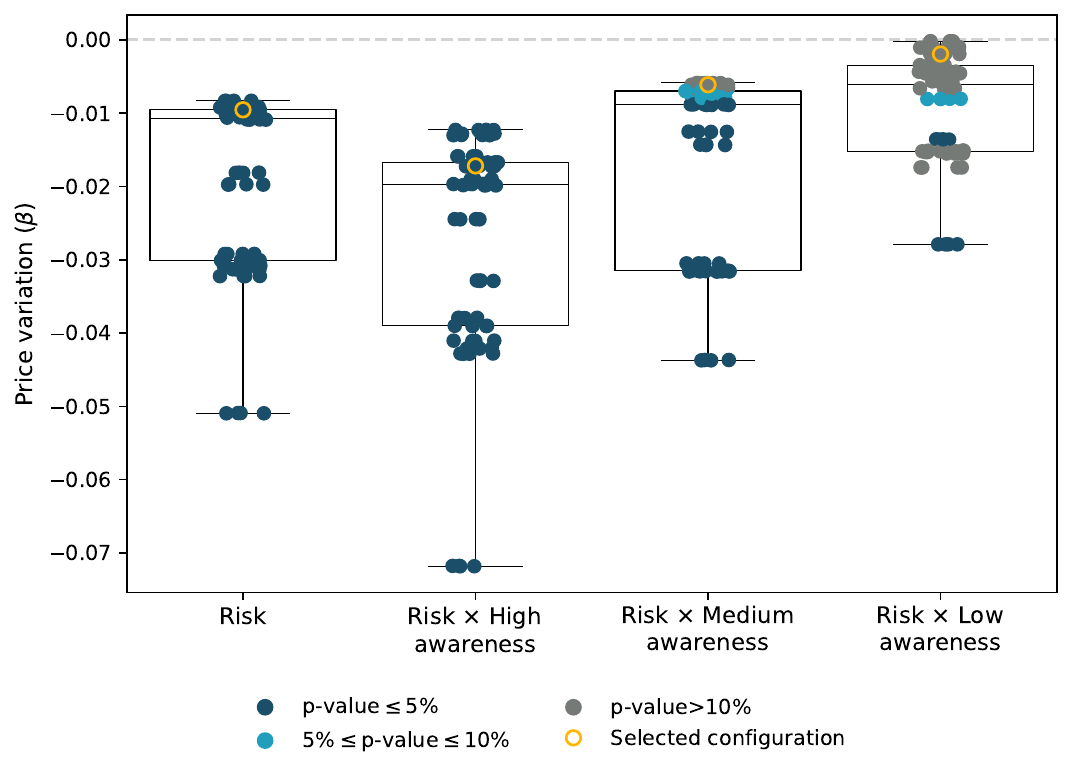}
    \caption{Effect on home prices of being in a flood-risk area (identified using ISPRA risk maps, first-left box) across Italy and considering high, medium, and low awareness transactions based on the flood events recorded by~\cite{emdat} and $\tau=10$ years, second to fourth boxes. Each dot represents a parameter configuration, the color reports the statistical significance. The dots with the yellow edges identify the parameter configuration presented in the main text. R$^2$ [min, max] = [0.48, 0.85]; observations [min, max] = [552761, 555880].}
    \label{fig:AllItaly_config}
\end{figure}

\begin{figure}
    \centering
    \includegraphics[width=1\linewidth]{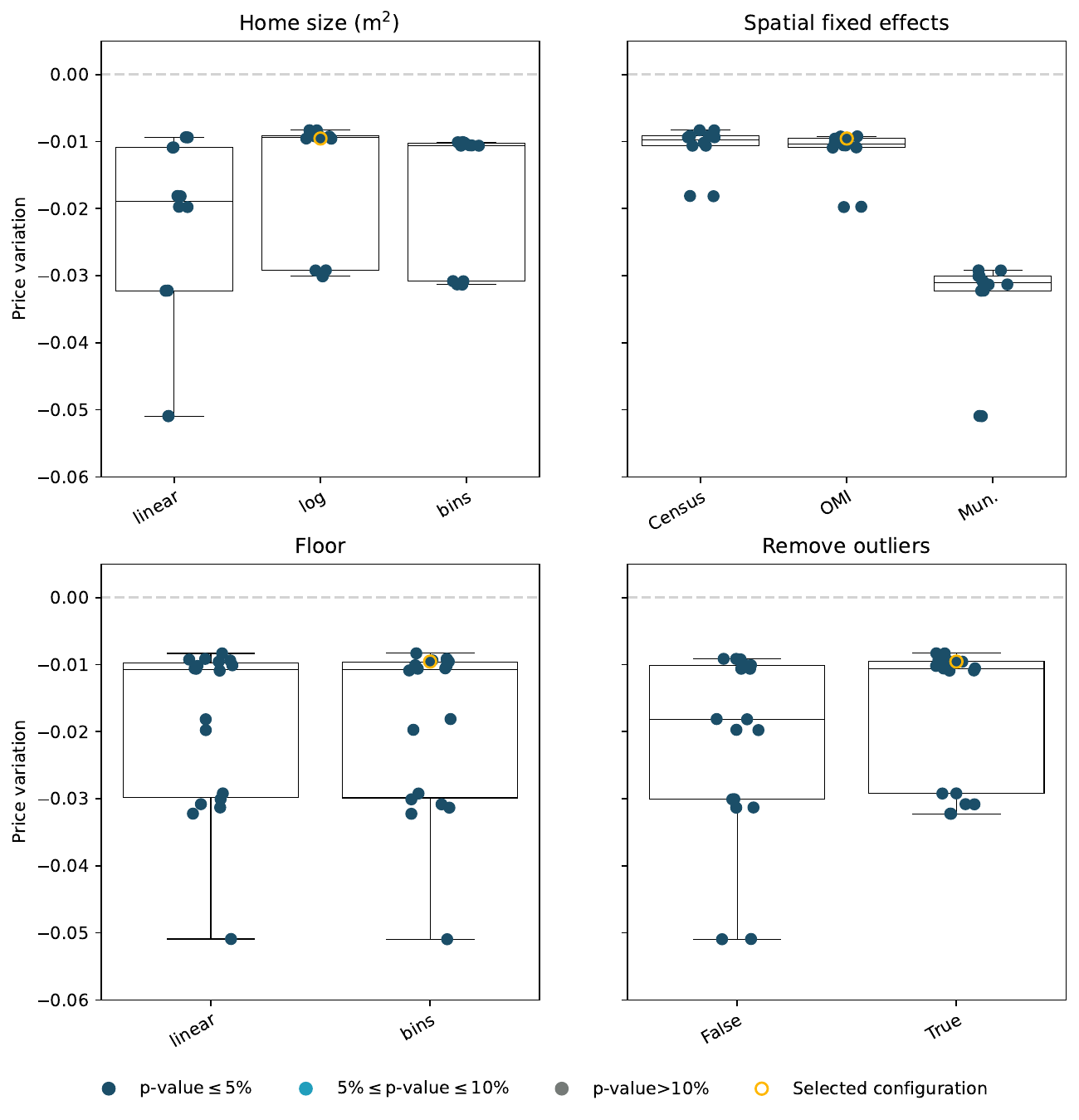}
    \caption{Changes of the home price across home size specifications --linear, logarithm (log) and bins of the square meters-- for homes in flood risk areas. Each dot represents a parameter configuration, the color reports the statistical significance. The dots with the yellow edges identify the parameter configuration presented in the main text. Each panel refers to a single variable reported in the panel title, e.g., the top-left panel describes how the home size specification affects the price variations for homes in flood-risk areas. R$^2$ [min, max] = [0.48,0.85]; observations [min, max] = [552761, 555880].}
    \label{fig:mq_config}
\end{figure}

Figure~\ref{fig:AllItaly_config} shows the effect on the home price of being in a risk zone across Italy for all parameter configurations and half memory time (see Section~\ref{sec:frequency}) of 10 years. Flood risk causes a statistically significant price reduction, ranging between 1\% and 5\%. Once we focus on transaction with high, medium, and low awareness, the largest price reductions are observed for the high awareness purchases. Instead, coefficients for the medium and low awareness transactions are generally smaller and less significant than those for high awareness. The yellow circle in the plot marks the specification selected for the main text, which considers: the logarithm of home size, binned floors, fixed effects at OMI microzone, and the outliers have been removed. We can observe that the results shown in the main are representative also of other parameter specifications since they are in the bulk of points. The exception is the chunk of significant negative coefficients for medium and low awareness transactions. These coefficients are associated with the municipality fixed effects as confirmed in Figure~\ref{fig:mq_config}, and so probably highlight other factors than flood risk.

Figure~\ref{fig:mq_config} details the price variation due to modifying individually each of the parameters explored. In all cases, the parameter configuration shown in the main results (highlighted in yellow) agrees with other specifications in terms of the coefficient's order of magnitude and its statistical significance. The spatial fixed effects at the municipality scale (top-right panel) present some differences. In this case, risk coefficients are lower than those associated with census track and OMI microzones. We argue that, when controlling for municipality, the effects of flood risk are mixed with those of peripheral home locations, leading to the mentioned discrepancies. 

For completeness, we provide the coefficients for all the non-risk related variables (except the spatial and temporal fixed effects) of our baseline risk specification (corresponding Model 3 in Table~\ref{tab:NewBaselinePriceTable_EMDAT_numflood}). The results are shown in Table\ref{tab:hedonic}.

\begin{small}
\begin{longtable}{llll}
\caption{Coefficients for all the variables in the baseline hedonic regression (third column of Table \ref{tab:NewBaselinePriceTable_EMDAT_numflood}).} \label{tab:hedonic} \\
\toprule
 & Coefficient & Std. Error & stars \\
\midrule
\endfirsthead
\caption[]{Hedonic coefficients for the first column of Table~\ref{tab:NewBaselinePriceTable_EMDAT_numflood}} \\
\toprule
 & Coefficient & Std. Error & stars \\
\midrule
\endhead
\midrule
\multicolumn{4}{r}{Continued on next page} \\
\midrule
\endfoot
\bottomrule
\endlastfoot
Cadastral code: A02 (ref.: A01) & -0.120 & 0.024 & *** \\
Cadastral code: A03 (ref.: A01) & -0.208 & 0.024 & *** \\
Cadastral code: A04 (ref.: A01) & -0.319 & 0.025 & *** \\
Cadastral code: A05 (ref.: A01) & -0.405 & 0.029 & *** \\
Cadastral code: A06 (ref.: A01) & -0.337 & 0.054 & *** \\
Cadastral code: A07 (ref.: A01) & -0.028 & 0.025 &  \\
Cadastral code: A08 (ref.: A01) & 0.174 & 0.047 & *** \\
Cadastral code: A09 (ref.: A01) & -0.184 & 0.286 &  \\
Cadastral code: A10 (ref.: A01) & -0.239 & 0.054 & *** \\
Cadastral code: A11 (ref.: A01) & -0.159 & 0.130 &  \\
Construction year: 1955-1960 (ref.: $<$1955) & 0.021 & 0.003 & *** \\
Construction year: 1960-1965 (ref.: $<$1955) & -0.000 & 0.004 &  \\
Construction year: 1965-1970 (ref.: $<$1955) & 0.045 & 0.003 & *** \\
Construction year: 1970-1975 (ref.: $<$1955) & 0.050 & 0.004 & *** \\
Construction year: 1975-1985 (ref.: $<$1955) & 0.084 & 0.004 & *** \\
Construction year: 1985-1995 (ref.: $<$1955) & 0.175 & 0.004 & *** \\
Construction year: 1995-2005 (ref.: $<$1955) & 0.270 & 0.004 & *** \\
Construction year: 2005-2015 (ref.: $<$1955) & 0.343 & 0.004 & *** \\
Construction year: 2015-2025 (ref.: $<$1955) & 0.404 & 0.006 & *** \\
Construction year: Missing (ref.: $<$1955) & 0.046 & 0.013 & *** \\
Energy class: A2 (ref.: A) & 0.005 & 0.006 &  \\
Energy class: A3 (ref.: A) & 0.027 & 0.008 & *** \\
Energy class: A4 (ref.: A) & 0.052 & 0.006 & *** \\
Energy class: B (ref.: A) & -0.014 & 0.005 & ** \\
Energy class: C (ref.: A) & -0.048 & 0.005 & *** \\
Energy class: D (ref.: A) & -0.062 & 0.005 & *** \\
Energy class: E (ref.: A) & -0.079 & 0.005 & *** \\
Energy class: F (ref.: A) & -0.110 & 0.006 & *** \\
Energy class: G (ref.: A) & -0.145 & 0.005 & *** \\
Energy class: Missing (ref.: A) & -0.100 & 0.006 & *** \\
Flag air conditioning: True (ref.: False) & 0.051 & 0.002 & *** \\
Flag air conditioning: Missing (ref.: False) & 0.009 & 0.002 & *** \\
Flag garage & 0.107 & 0.002 & *** \\
Flag multi floor: True & 0.015 & 0.004 & *** \\
Flag annex & 0.018 & 0.002 & *** \\
Floor: 1 (ref.: 0) & -0.027 & 0.003 & *** \\
Floor: 2 (ref.: 0) & -0.028 & 0.003 & *** \\
Floor: 3 (ref.: 0) & -0.026 & 0.004 & *** \\
Floor: 4+ (ref.: 0) & 0.008 & 0.005 &  \\
Floor: Missing (ref.: 0) & -0.004 & 0.003 &  \\
Logarithm of square meters & 0.706 & 0.003 & *** \\
\end{longtable}
\end{small}

\subsubsection*{Robustness historical memory definition}

To validate our definition of the historical memory of regions we repeated the hedonic regression analyses in Table~\ref{tab:NewBaselinePriceTable_EMDAT_numflood} after considering an alternative dataset for past flood events that we compiled from ISPRA records.

EM-DAT is well-known to over-represent disasters in richer areas \citep{felbermayr2014naturally}. Therefore, we compiled a dataset of flood events from ISPRA data and re-estimated our hedonic regressions. ISPRA collects data on flood events drawing on multiple sources, such as the Italian Civil Protection Department, ISTAT, the National Research Council, local administrations and newspapers. For each event, it reports the duration, an estimation of the damages, an estimation of the allocated aid funds, and the total number of deaths and missing or evacuated people. Although data are provided back to 1951, systematic data collection only began in 2003, so we only consider the later period for better accuracy. Data are not collected in tabular format, but we manually compiled a database reading from the yearly reports by ISPRA.\footnote{\url{https://indicatoriambientali.isprambiente.it/it/edizioni-annuario}. The latest available report is for the year 2022.} The ISPRA dataset contains 402 event-region pairs distributed across regions and macroareas. Since the inclusion criteria for ISPRA are less strict than EM-DAT, this results on some variations in the historical memory classifications. Even with these differences we still observe the same pattern of significance of the risk coefficients disaggregated by awareness conditions that was observed in the regression presented in the main manuscript (Table~\ref{tab:baselineRegr_ISPRA_numflood}).

\begin{table}[]
\caption{Flood risk effect on home prices disaggregated across terciles of awareness at the time of the transaction for different spatial fixed effects and half life times ($\tau$, see Section~\ref{sec:frequency}). $^{***} p < $ 0.01, $^{**} p < $ 0.05, and $^{*} p < $ 0.1. The awareness values are computed considering the flood events collected by~\citep{emdat}. }
\label{tab:SM_table2}

\rotatebox{90}{

\begin{tabular}{l|ccccccccc}
\hline
                                                                                   & \multicolumn{9}{c}{Dep. var.: log(\textit{P})}                                                                                                                                                                                                                                                                                                                                                                                                                                                                                                                                                                                                   \\ \hline
\textbf{Model id}                                                                  & 1                                                                & 2                                                                & \multicolumn{1}{c|}{3}                                                                & 4                                                                & 5                                                                & \multicolumn{1}{c|}{6}                                                                & 7                                                                & 8                                                                & 9                                                               \\ \hline
\textbf{Half time memory ($\tau$)}                                                 & 7                                                                & 10                                                               & \multicolumn{1}{c|}{17}                                                               & 7                                                                & 10                                                               & \multicolumn{1}{c|}{17}                                                               & 7                                                                & 10                                                               & 17                                                              \\[10pt]
\textbf{\begin{tabular}[c]{@{}l@{}}Risk $\times$ High \\ awareness\end{tabular}}   & \begin{tabular}[c]{@{}c@{}}-0.040$^{**}$\\ (0.016)\end{tabular}  & \begin{tabular}[c]{@{}c@{}}-0.038$^{**}$\\ (0.016)\end{tabular}  & \multicolumn{1}{c|}{\begin{tabular}[c]{@{}c@{}}-0.032$^{**}$\\ (0.015)\end{tabular}}  & \begin{tabular}[c]{@{}c@{}}-0.015$^{***}$\\ (0.004)\end{tabular} & \begin{tabular}[c]{@{}c@{}}-0.017$^{***}$\\ (0.004)\end{tabular} & \multicolumn{1}{c|}{\begin{tabular}[c]{@{}c@{}}-0.014$^{***}$\\ (0.005)\end{tabular}} & \begin{tabular}[c]{@{}c@{}}-0.011$^{***}$\\ (0.004)\end{tabular} & \begin{tabular}[c]{@{}c@{}}-0.012$^{***}$\\ (0.004)\end{tabular} & \begin{tabular}[c]{@{}c@{}}-0.009$^{**}$\\ (0.004)\end{tabular} \\[10pt]
\textbf{\begin{tabular}[c]{@{}l@{}}Risk $\times$ Medium \\ awareness\end{tabular}} & \begin{tabular}[c]{@{}c@{}}-0.029$^{***}$\\ (0.008)\end{tabular} & \begin{tabular}[c]{@{}c@{}}-0.031$^{***}$\\ (0.009)\end{tabular} & \multicolumn{1}{c|}{\begin{tabular}[c]{@{}c@{}}-0.036$^{***}$\\ (0.009)\end{tabular}} & \begin{tabular}[c]{@{}c@{}}-0.007$^{*}$\\ (0.004)\end{tabular}   & \begin{tabular}[c]{@{}c@{}}-0.007$^{*}$\\ (0.004)\end{tabular}   & \multicolumn{1}{c|}{\begin{tabular}[c]{@{}c@{}}-0.010$^{**}$\\ (0.004)\end{tabular}}  & \begin{tabular}[c]{@{}c@{}}-0.007$^{}$\\ (0.004)\end{tabular}    & \begin{tabular}[c]{@{}c@{}}-0.008$^{*}$\\ (0.004)\end{tabular}   & \begin{tabular}[c]{@{}c@{}}-0.010$^{**}$\\ (0.004)\end{tabular} \\[10pt]
\textbf{\begin{tabular}[c]{@{}l@{}}Risk $\times$ Low \\ awareness\end{tabular}}    & \begin{tabular}[c]{@{}c@{}}-0.014$^{}$\\ (0.012)\end{tabular}    & \begin{tabular}[c]{@{}c@{}}-0.014$^{}$\\ (0.013)\end{tabular}    & \multicolumn{1}{c|}{\begin{tabular}[c]{@{}c@{}}-0.018$^{}$\\ (0.013)\end{tabular}}    & \begin{tabular}[c]{@{}c@{}}-0.004$^{}$\\ (0.004)\end{tabular}    & \begin{tabular}[c]{@{}c@{}}-0.002$^{}$\\ (0.004)\end{tabular}    & \multicolumn{1}{c|}{\begin{tabular}[c]{@{}c@{}}-0.002$^{}$\\ (0.004)\end{tabular}}    & \begin{tabular}[c]{@{}c@{}}-0.006$^{}$\\ (0.004)\end{tabular}    & \begin{tabular}[c]{@{}c@{}}-0.005$^{}$\\ (0.004)\end{tabular}    & \begin{tabular}[c]{@{}c@{}}-0.006$^{}$\\ (0.004)\end{tabular}   \\[10pt] \hline
\textbf{Observations}                                                              & \multicolumn{3}{c|}{552728}                                                                                                                                                                                                 & \multicolumn{3}{c|}{552728}                                                                                                                                                                                                 & \multicolumn{3}{c}{552728}                                                                                                                                                                            \\
\textbf{R\textsuperscript{2}}                                     & \multicolumn{3}{c|}{0.669}                                                                                                                                                                                                  & \multicolumn{3}{c|}{0.755}                                                                                                                                                                                                  & \multicolumn{3}{c}{0.848}                                                                                                                                                                             \\ \hline
\textbf{Spatial fix. effects}                                                      & \multicolumn{3}{c|}{Municipality}                                                                                                                                                                                           & \multicolumn{3}{c|}{OMI zone}                                                                                                                                                                                               & \multicolumn{3}{c}{Census Tract}                                                                                                                                                                      \\
\textbf{N. clusters}                                                               & \multicolumn{3}{c|}{7203}                                                                                                                                                                                                   & \multicolumn{3}{c|}{20297}                                                                                                                                                                                                  & \multicolumn{3}{c}{160630}                                                                                                                                                                            \\
\textbf{Clustered s.e.}                                                            & \multicolumn{3}{c|}{\checkmark}                                                                                                                                                                              & \multicolumn{3}{c|}{\checkmark}                                                                                                                                                                              & \multicolumn{3}{c}{\checkmark}                                                                                                                                                         \\ \hline
\end{tabular}

}
\end{table}

\begin{table}[]
\centering
\caption{Flood risk effect on home prices. For each spatial fixed effects, Models 1-3-5 report the effect of the risk on home prices estimated using Equation~\eqref{eq:baseline}, while Models 2-4-6 disaggregate the effect of flood risk by the level of flood awareness at the time of the purchase, Equation~\eqref{eq:hist_mem}. Awareness is estimated following the procedure described in~\ref{sec:frequency} using the data reported by ISPRA and $\tau=10$ years. *** $p < 0.01$, ** $p < 0.05$, * $p < 0.1$.}
\begin{tabular}{l|cccccc}
\hline
                                                                                   & \multicolumn{6}{c}{Dep. var.: log(\textit{P})}                                                                                                                                                                                                                                                                                                                                                                                          \\ \hline
\textbf{Model id}                                                                  & 1                                                                & \multicolumn{1}{c|}{2}                                                                & 3                                                                & \multicolumn{1}{c|}{4}                                                                & 5                                                               & 6                                                                \\ \hline
\textbf{Risk}                                                                      & \begin{tabular}[c]{@{}c@{}}-0.029$^{***}$\\ (0.010)\end{tabular} & \multicolumn{1}{c|}{}                                                                 & \begin{tabular}[c]{@{}c@{}}-0.010$^{***}$\\ (0.003)\end{tabular} & \multicolumn{1}{c|}{}                                                                 & \begin{tabular}[c]{@{}c@{}}-0.008$^{**}$\\ (0.003)\end{tabular} &                                                                  \\[10pt]
\textbf{\begin{tabular}[c]{@{}l@{}}Risk $\times$ High \\ awareness\end{tabular}}   &                                                                  & \multicolumn{1}{c|}{\begin{tabular}[c]{@{}c@{}}-0.039$^{**}$\\ (0.017)\end{tabular}}  &                                                                  & \multicolumn{1}{c|}{\begin{tabular}[c]{@{}c@{}}-0.019$^{***}$\\ (0.005)\end{tabular}} &                                                                 & \begin{tabular}[c]{@{}c@{}}-0.015$^{***}$\\ (0.005)\end{tabular} \\[10pt]
\textbf{\begin{tabular}[c]{@{}l@{}}Risk $\times$ Medium \\ awareness\end{tabular}} &                                                                  & \multicolumn{1}{c|}{\begin{tabular}[c]{@{}c@{}}-0.029$^{***}$\\ (0.011)\end{tabular}} &                                                                  & \multicolumn{1}{c|}{\begin{tabular}[c]{@{}c@{}}-0.004$^{}$\\ (0.004)\end{tabular}}    &                                                                 & \begin{tabular}[c]{@{}c@{}}-0.007$^{}$\\ (0.004)\end{tabular}    \\[10pt]
\textbf{\begin{tabular}[c]{@{}l@{}}Risk $\times$ Low \\ awareness\end{tabular}}    &                                                                  & \multicolumn{1}{c|}{\begin{tabular}[c]{@{}c@{}}-0.015$^{}$\\ (0.014)\end{tabular}}    &                                                                  & \multicolumn{1}{c|}{\begin{tabular}[c]{@{}c@{}}-0.001$^{}$\\ (0.004)\end{tabular}}    &                                                                 & \begin{tabular}[c]{@{}c@{}}-0.001$^{}$\\ (0.005)\end{tabular}    \\ \hline
\textbf{Observations}                                                              & \multicolumn{2}{c|}{552728}                                                                                                                              & \multicolumn{2}{c|}{552728}                                                                                                                              & \multicolumn{2}{c}{552728}                                                                                                         \\
\textbf{R\textsuperscript{2}}                                     & \multicolumn{2}{c|}{0.669}                                                                                                                               & \multicolumn{2}{c|}{0.755}                                                                                                                               & \multicolumn{2}{c}{0.848}                                                                                                          \\ \hline
\textbf{Spatial fix. effects}                                                      & \multicolumn{2}{c|}{Municipality}                                                                                                                        & \multicolumn{2}{c|}{OMI zone}                                                                                                                            & \multicolumn{2}{c}{Census Tract}                                                                                                   \\
\textbf{N. clusters}                                                               & \multicolumn{2}{c|}{7203}                                                                                                                                & \multicolumn{2}{c|}{20297}                                                                                                                               & \multicolumn{2}{c}{160630}                                                                                                         \\
\textbf{Clustered s.e.}                                                            & \multicolumn{2}{c|}{\checkmark}                                                                                                           & \multicolumn{2}{c|}{\checkmark}                                                                                                           & \multicolumn{2}{c}{\checkmark}                                                                                      \\ \hline
\end{tabular}
\label{tab:baselineRegr_ISPRA_numflood}
\end{table}

\end{document}